\documentclass{aa}   

\usepackage{graphicx}
\usepackage{txfonts}
\usepackage{natbib,twoopt}
\usepackage{lscape}
\usepackage[breaklinks=true]{hyperref} 
\bibpunct{(}{)}{;}{a}{}{,} 
\makeatletter
\newcommandtwoopt{\citeads}[3][][]{\href{http://adsabs.harvard.edu/abs/#3}%
{\def\hyper@linkstart##1##2{}%
\let\hyper@linkend\@empty\citealp[#1][#2]{#3}}}
\newcommandtwoopt{\citepads}[3][][]{\href{http://adsabs.harvard.edu/abs/#3}%
{\def\hyper@linkstart##1##2{}%
\let\hyper@linkend\@empty\citep[#1][#2]{#3}}}
\newcommandtwoopt{\citetads}[3][][]{\href{http://adsabs.harvard.edu/abs/#3}%
{\def\hyper@linkstart##1##2{}%
\let\hyper@linkend\@empty\citet[#1][#2]{#3}}}
\newcommandtwoopt{\citeyearads}[3][][]%
{\href{http://adsabs.harvard.edu/abs/#3}
{\def\hyper@linkstart##1##2{}%
\let\hyper@linkend\@empty\citeyear[#1][#2]{#3}}}
\makeatother

\begin{document}

   \title{Multiplicity among solar-type stars}

   \subtitle{IV. The CORAVEL radial velocities and the spectroscopic orbits of
nearby K dwarfs\thanks{Based on photoelectric radial-velocity measurements collected 
at Haute-Provence Observatory.}}

   \author{J.-L. Halbwachs
          \inst{1}
          \and
          M. Mayor\inst{2}
          \and
          S. Udry\inst{2}
          }

   \institute{Universit\'e de Strasbourg, CNRS,
   Observatoire Astronomique de Strasbourg, UMR 7550,
   11, rue de l'Universit\'{e},
   F--67\,000 Strasbourg, France \\
   \email{jean-louis.halbwachs@astro.unistra.fr}
         \and
             Observatoire Astronomique de l'Universit\'e de Gen\`eve, 51, chemin des Maillettes,
             CH--1290 Sauverny, Switzerland \\
             }

   \date{Received 7 May 2018; accepted 2 August 2018}

 
  \abstract
   {The statistical properties of binary stars are clues for understanding their formation
    process. A radial velocity survey was carried on amongst nearby G-type stars and
    the results were published in 1991.} 
   {The survey of radial velocity measurements was extended towards K-type stars.}
   {A sample of 261 K-type stars was observed with the spectrovelocimeter CORAVEL (COrrelation RAdial VELocities). 
   Those stars with a variable radial velocity were detected on the basis of the $P(\chi^2)$ test.
   The orbital elements of the spectroscopic binaries were then derived.}
   {The statistical properties of binary stars were derived from these observations and published in 2003.
    We present the catalogue of the radial velocity measurements obtained with
    CORAVEL for all the K stars of the survey and the orbital elements 
    derived for 34 spectroscopic systems. In addition, the catalogue contains eight G-type spectroscopic binaries
    that have received additional measurements since 1991 and for which the orbital elements
    are revised or derived for the first time. 
    }
   {}


   \maketitle
%

\section{Introduction}
\label{introduction}

The spectrovelocimeter CORAVEL \citep[COrrelation RAdial VELocities,][]{Baranne79} 
was installed on the Swiss 1-m telescope at the Observatory of Haute-Provence (OHP) from
the late 1970s until its decommissioning in 2000. Amongst other programmes, it provided
the radial-velocity (RV) measurements exploited in two statistical studies of binarity 
among the stars in the solar neighbourhood: the study of solar--type stars until G8,
and its extension towards the K-type stars. A series of articles has been devoted to
these programmes. The first \citep[][Paper~I hereafter]{DMH91} presented the radial-velocity measurements
of the sample of F-G type stars; these data led to the orbital
elements of several spectroscopic binaries (SBs), and to the statistical properties of
solar-type binaries \citep[][DM91 hereafter]{DM91}. Later, \citet[][Paper~III hereafter]{Halbwachs03}
extended the statistical investigations to the K-type binaries with periods shorter than
ten years, again on the basis of
CORAVEL observations. This paper presented the parameters relevant for statistics, namely
the periods, the semi-amplitudes, the mass ratios, and the orbital eccentricities of the
spectroscopic binaries, excluding the other orbital elements. The long period K-type binaries
were eventually studied by \citet{Eggenberger04}.

\defcitealias{DMH91}{Paper~I}
\defcitealias{DM91}{DM91}
\defcitealias{Halbwachs03}{Paper~III}

The present paper completes the series by presenting the radial velocity measurements and the
full set of orbital elements that gave rise to \citetalias{Halbwachs03}. 
It will give the orbits we have discovered all the visibility they deserve, so that they are
henceforth taken into account in statistical studies, such as that of \citet{Raghavan}.
Moreover, they will be available for the validation of the spectroscopic orbits derived
from the Radial Velocity Spectrometer of the Gaia satellite \citep{GaiaRVS}.
The CORAVEL programme
is presented in Sect.~\ref{CORAVEL}, the RV catalogue is in Sect.~\ref{RV}, and the 
spectroscopic orbital elements are in Sect.~\ref{orbits}. Section~\ref{conclusion} is the
conclusion.

\section{The CORAVEL survey of nearby K-type stars}
\label{CORAVEL}

The CORAVEL survey for nearby SBs was initiated in the early 1980s, although
some stars (especially among the F--G types) had been measured before. The stars
were taken from the second edition of the ``Catalogue of Nearby Stars'' 
\citep[CNS][]{Gliese} and from its supplement \citep{GlieJahr}. 
The stars discarded from the preliminary third version
of the CNS \citep[CNS3][]{GJ91} were kept in the observing runs. All stars were
observed with CORAVEL from the
Haute--Provence Observatory. Due to the location of the instrument and to
its characteristics, only the stars as late as F7 and northern to $-15\ ^{\circ}$ in
declination were observed. Some stars with declination below  $-15\ ^{\circ}$ were
observed, but they were not taken into account in \citetalias{Halbwachs03}.
The programme was split according
to the spectral types of the stars: the search for SBs amongst 288 F--G stars ended in
December 1989, but the detection of SBs amongst the K--type stars was intensively 
performed until July 1993. After this date, the SBs were observed until
2000; at the same time, the RV of a few stars were still measured in order 
to confirm that it was constant.

\section{Radial-velocity catalogues}
\label{RV}

\subsection{CORAVEL individual measurements}
\label{RV:CORAVEL}

The catalogue of the RV measurements provides 5413 measurements for 269 stars:
261 K--type stars and eight stars from the sample of \citetalias{DM91}. These
eight G--type stars were already in \citetalias{DMH91} or another paper quoted in
\citetalias{DM91}, but they fulfil two
conditions: they received enough additional RV measurements
between 1991 and 2000 to significantly improve their spectroscopic orbit, and
this new CORAVEL orbit was not published elsewhere.
Moreover, the reduction of the CORAVEL observations
was slightly improved, and the RV measurements are not exactly the same as in
\citetalias{DMH91}.

\begin{table*}
\caption{Sample records of the RV measurement catalogues. The first records refer to
CORAVEL RVs, and the last records refer to the file of the Elodie RV measurements.}             
\label{table:RV}      
\centering                          
\begin{tabular}{rccrrcc}        
\hline\hline                 
\multicolumn{1}{c}{GJ} & \multicolumn{2}{c}{epoch\phantom{BJD0}}&    RV\phantom{04}  &$\sigma_{RV}\phantom{04}$&comp.& remark \\
       &yymmdd&BJD$-2\,400\,000$&km~s$^{-1}$&  km~s$^{-1}$& &      \\
\hline   
   5\phantom{.1A}&780213&43553.279\phantom{2}&   $-5.90$\phantom{3}& 0.41\phantom{4} &1&    \\  
   5\phantom{.1A}&780825&43746.491\phantom{2}&   $-7.00$\phantom{3}& 0.32\phantom{4} &1&    \\  
   5\phantom{.1A}&780914&43766.509\phantom{2}&   $-6.08$\phantom{3}& 0.35\phantom{4} &1&    \\  
   5\phantom{.1A}&830903&45581.567\phantom{2}&   $-5.92$\phantom{3}& 0.33\phantom{4} &1&    \\  
   5\phantom{.1A}&851219&46419.286\phantom{2}&   $-6.52$\phantom{3}& 0.31\phantom{4} &1&    \\    
  ..\phantom{.1A}&..    &..\phantom{2}       &   ..\phantom{3}   & ..\phantom{4}   &.&    \\      
  53.1A&771103&43451.441\phantom{2}&    6.94\phantom{3}& 0.32\phantom{4} &1&    \\ 
  ..\phantom{.1A}&..    &..\phantom{2}       &   ..\phantom{3}   & ..\phantom{4}   &.&    \\  
 1124\phantom{.1A}&870330&46885.380\phantom{2}&  $-54.20$\phantom{3}& 0.89\phantom{4} &2& R  \\
\hline 
1069\phantom{.1A}&970821&50681.6321          & 24.440& 0.044&1&    \\  
  ..\phantom{.1A}&..    &..\phantom{2}       &   ..\phantom{3}   & ..\phantom{4}   &.&    \\  
 554\phantom{.1A}&980109&50822.7302          &$-18.863$& 0.014&1& R  \\
\hline                                   
\end{tabular}
\end{table*}

The format of the catalogue is presented in Table~\ref{table:RV}.
Each record consists in the following data:

\begin{itemize}
\item
The number of the star in the CNS, followed by a letter designating the
component, if any.
\item
The epoch of the observation, given as a date with the year, the month, and the day,
and also as barycentric Julian Day (BJD). 
\item
The RV, in km~s$^{-1}$.
\item
The uncertainty of the RV.
\item
The index of component (``1'' for the primary, ``2'' for the secondary).
\item
A flag ``R'' indicates the measurement was discarded from the calculation of
the orbital elements.
\end{itemize}

The records are sorted by stars (from the smallest to the largest right ascension),
and then by observation epochs.

\subsection{Elodie individual measurements}
\label{RV:Elodie}

The CORAVEL observations were not sufficiently accurate to derive valuable SB
orbital elements for two stars of the sample, GJ 1069 and GJ 554. The latter
of these two stars is even a constant velocity star when only the CORAVEL
RVs are considered. Fortunately, RV measurements were performed thanks to the
spectrograph Elodie of the 193 cm telescope at the Haute-Provence Observatory,
and they are provided by the Elodie archive (http://atlas.obs-hp.fr/elodie/).
Fifteen Elodie RVs of GJ 1069 and 58 RVs of GJ 554 are presented in a separate file,
with a slightly different format due to their accuracy. The uncertainties
of the Elodie RVs were estimated as explained in Sect.~\ref{orbitElodie}.
Sample records are presented at the end of Table~\ref{table:RV}.

\subsection{The mean RV and the detection of SBs}

The RV measurements were used to derive statistical information to decide whether a 
star is binary or not. These data are provided in Table~\ref{table:meanRV}.

\begin{figure*}
\includegraphics[clip=,height=20.5 cm]{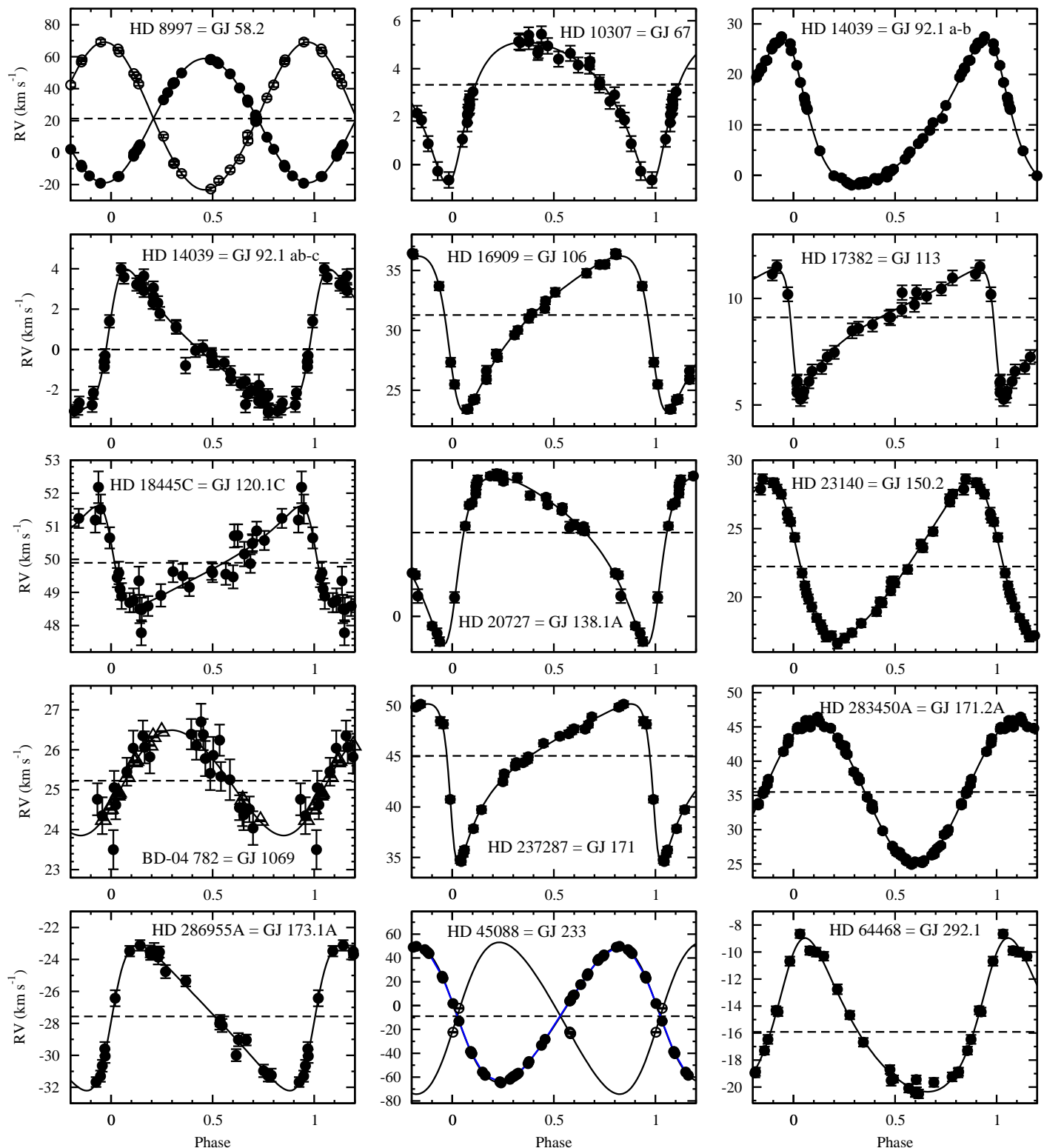} 
 \caption{Spectroscopic orbits of the first 15 SBs in Table~\ref{tab:orbites}; the circles refer to the 
non-rejected CORAVEL RV measurements and, for GJ 1069, the open
triangles refer to the measurements obtained with Elodie; the Elodie RVs are shifted to the
zero point of the CORAVEL measurements.}
\label{fig:orb1}
\end{figure*}

\begin{figure*}
\includegraphics[clip=,height=20.5 cm]{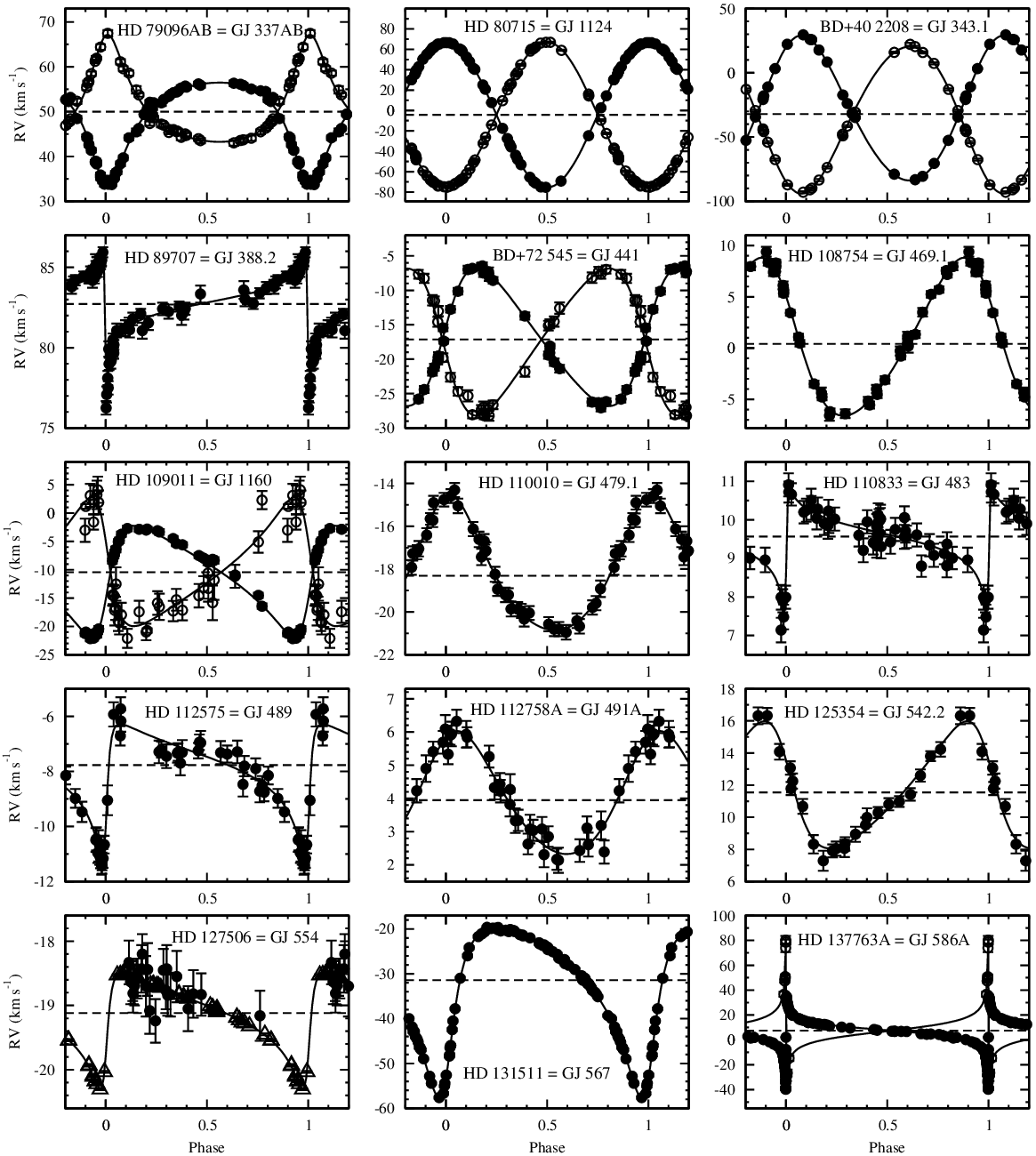} 
 \caption{Spectroscopic orbits of the second set of 15 SBs in Table~\ref{tab:orbites}; the circles refer to the 
non-rejected CORAVEL RV measurements and, for GJ 554, the open
triangles refer to the measurements obtained with Elodie; the Elodie RVs are shifted to the
zero point of the CORAVEL measurements.}
\label{fig:orb2}
\end{figure*}

\begin{figure*}
\includegraphics[clip=,height=20.5 cm]{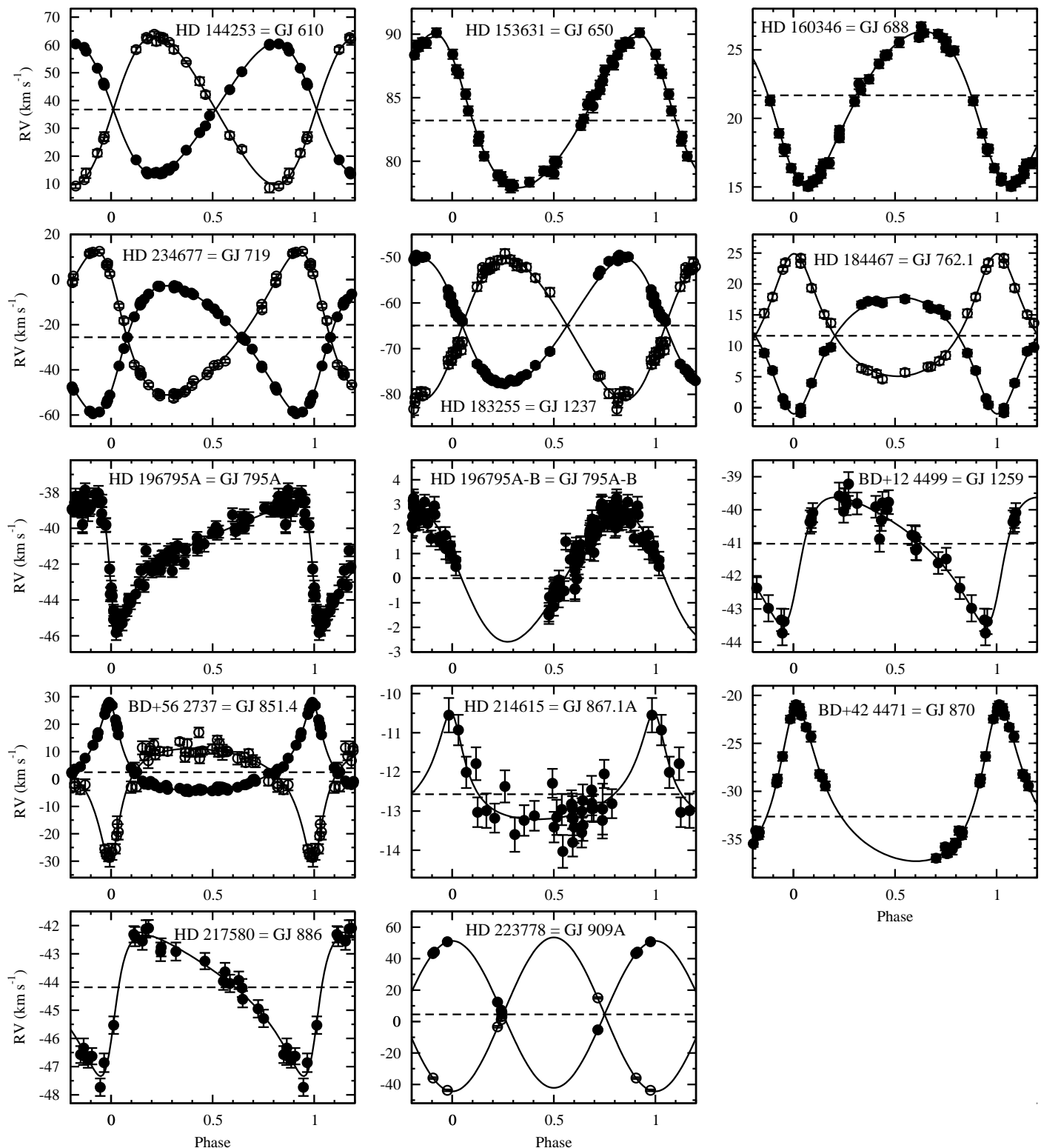} 
 \caption{Spectroscopic orbits of the 14 last SBs in Table~\ref{tab:orbites}; the circles refer to the 
non-rejected CORAVEL RV measurements.}
\label{fig:orb3}
\end{figure*}

The content is the following:

\begin{itemize}
\item
The identification of the star is the CNS number (GJ), as in the RV catalogue, and another
identification, which is HD when it exists, otherwise BD, or HIP, or AG (Astronomische Gesellschaft
catalogue). Three stars are designated
only by the GJ identification; these stars are all visual secondary components, as indicated by
letter ``B'' following their GJ number: GJ 57.1B, GJ 615.1B, and GJ 764.1B.
\item
The $B-V$ colour index used to derive the CORAVEL RVs of the star. For the eight stars from the G--type
sample, $B-V=0.63$ was assumed.
\item
The mean RV, $\overline{RV}$. When the star is a binary with known orbital elements, the RV of the barycentre is provided, as it 
is in Table~\ref{tab:orbites}.
\item
$\epsilon$ is the uncertainty of $\overline{RV}$.
\item
$\sigma_{RV}$ is the standard deviation of the RV measurements.
\item
$E/I$ is the ratio of external to internal errors.
\item
$P(\chi^2)$ is the probability to obtain the $\chi^2$ of the RVs of the star, assuming
that the RV is constant in reality.
\item
$N_1$ is the number of observations of the star.
\item
$N_{tot}$ is the number of RV measurements of the primary and of the secondary components.
\item
$\Delta T$ is the time span of the observation.
\item
The spectroscopic status is ``CST'' (ConStanT) when $P(\chi^2)$ is more than 1\%. Otherwise, it is
SB1, or SB2 when the RV of the secondary component was measured. An ``O'' indicates that the
orbital elements were derived, as explained in Sect.~\ref{orbits} hereafter.
One star, GJ~554, has a constant CORAVEL RV, but an SB orbit derived by adding
Elodie RV; this star is flagged ``CSTO''.
The status of two stars (GJ~893.2B and GJ~907.1) is ``?''
since only one RV measurement was obtained; these stars have both declinations around $-10\ ^{\circ}$,
and are too faint to be easily observable. However, they were only in the so-called ``extended
sample'' of \citetalias{Halbwachs03}, and they were not relevant in the statistical investigations.
\item
A flag indicates that the SB orbital elements of the star are in Table~\ref{tab:orbites}, ``*'' when
they are derived for the first time, and ``+'' otherwise. The velocity curves of the flagged stars
are in Figs.~\ref{fig:orb1}, \ref{fig:orb2}, and \ref{fig:orb3}.
\end{itemize}

\begin{figure}
\centering
   \includegraphics[bb=35 26 707 531,width=8.7cm,clip]{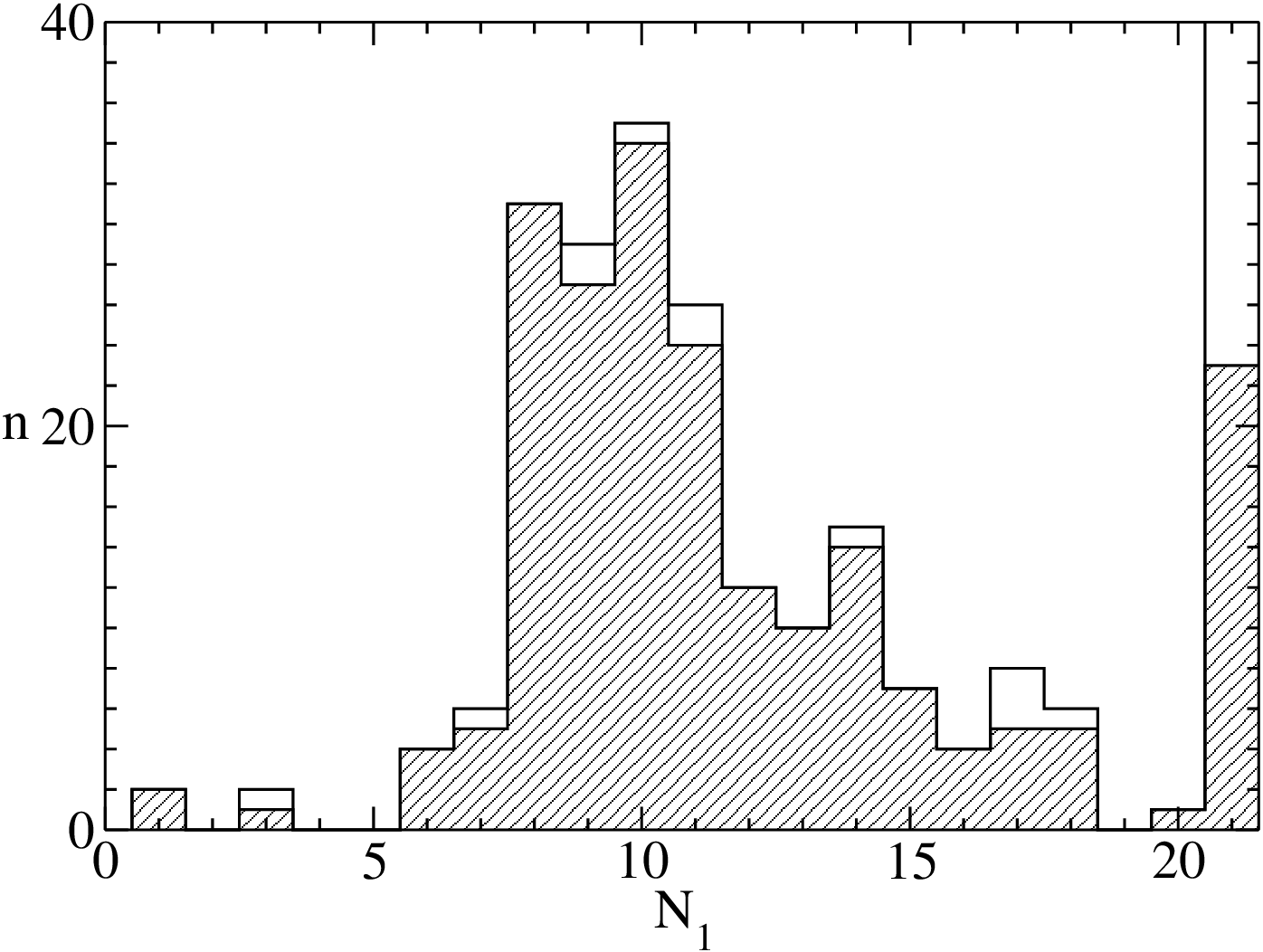}
   \caption{Histogram of the number of observations. The shaded area refers to the stars
that were not considered as variable and the white area to the spectroscopic binaries.
The last bin represents all the stars with at least 21 observations; for the spectroscopic
binaries, the count in this bin is 72.
           }
      \label{fig:HistoN1}
\end{figure}

\begin{figure}
\centering
   \includegraphics[bb=35 37 707 531,width=8.7cm,clip]{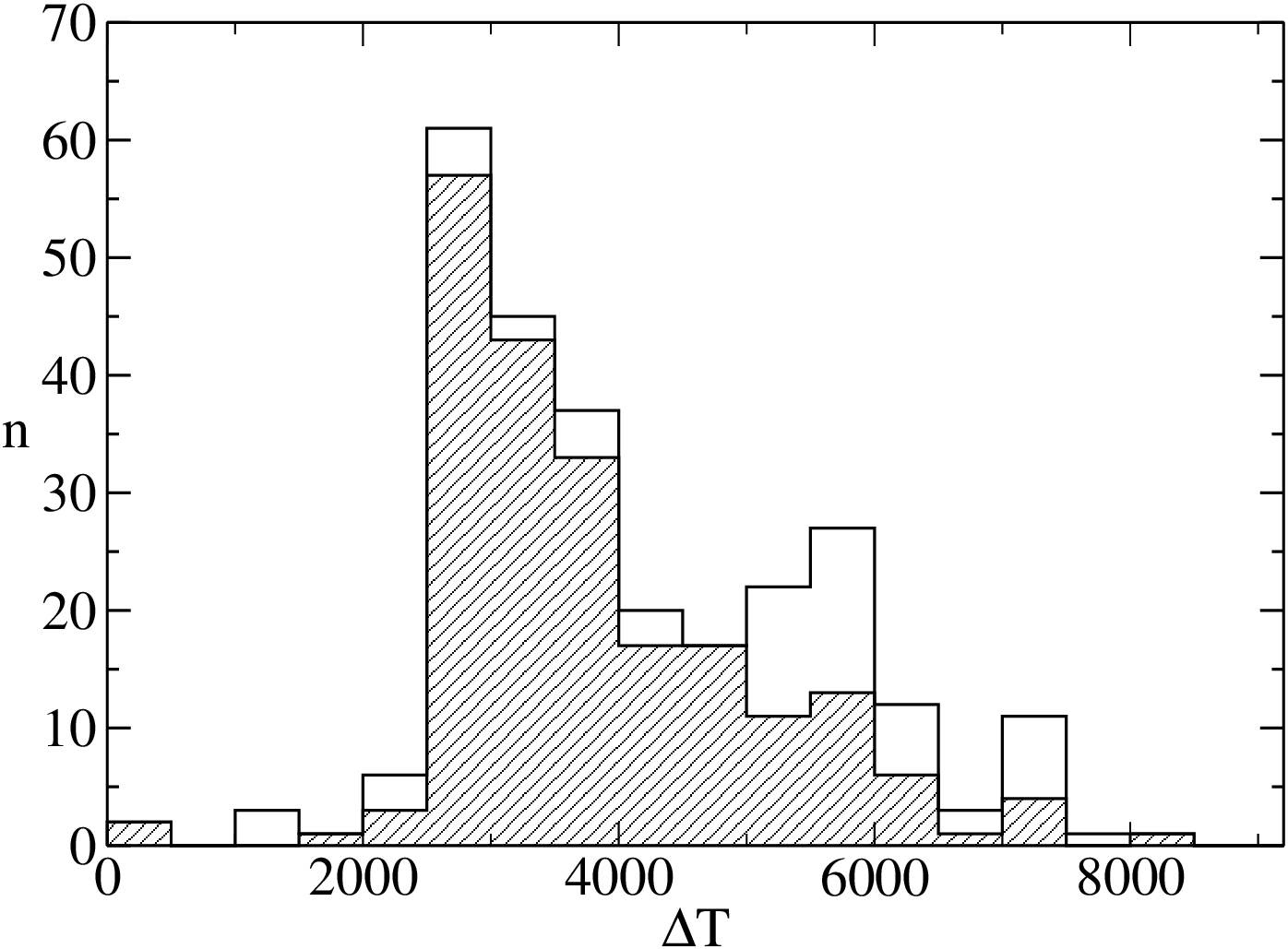}
   \caption{Histogram of the time span of the observations. The shaded area refers to the stars
that were not considered as variable and the white area to the spectroscopic binaries.
           }
      \label{fig:HistoDT}
\end{figure}

The efficiency of the detection of the SBs depends on the time spans of the RV survey and on the
numbers of RV measurements per star. The histograms of $N_1$ and of $\Delta T$ are presented in
Figs.~\ref{fig:HistoN1} and~\ref{fig:HistoDT}, respectively. The stars with variable RV are
counted apart from the others, since they received more observations when their variability 
was detected.

It appears from Fig.~\ref{fig:HistoN1} that a few stars received much less observations than the others.
In addition to the two stars with only one observation already mentioned above, two stars have
three observations, although one of them have a variable RV. This star is GJ~142, which was not
taken into account in the binarity statistics because its declination is close to $-20\ ^{\circ}$.
The other star is GJ~764.1B; it is difficult to observe since it is 5~arcsec away from its brighter companion
GJ~764.1A, and it is only in the ``extended'' sample. 

Half of the 209 constant RV stars received 11 observations or less. For the 269 stars in Table~\ref{table:meanRV},
the median number of observations is 12.
 
The distribution of the time span, Fig.~\ref{fig:HistoDT}, also indicates that a few stars seem
to have been less well observed than others. In addition to the two stars with one measurement and
$\Delta T=0$, 3 stars were observed during less than 1500 days, although their RVs were variable.
In fact, these stars are GJ~1124, GJ~343.1, and GJ~870, three short-period binaries, and their observations 
were completed in a few years; they received enough RV measurements to derive their orbital elements,
which are listed in Table~\ref{tab:orbites}. The median
time span is 3689 days for all the 270 stars, and 3410 days for the 209 constant RV stars.

\section{Orbital elements of the spectroscopic binaries}
\label{orbits}

\subsection{Taking into account the Elodie RV measurements of GJ~1069 and GJ~554}
\label{orbitElodie}
The accuracy of the CORAVEL RVs precludes the derivation of relevant orbital elements for GJ~1069 and
GJ~554, and it was necessary to take into account RVs provided by the Elodie archive. 
For that purpose, the uncertainty of the Elodie RVs must be estimated in order to assign them
a reliable weight with respect to the CORAVEL RVs. For each star, the same weight was assigned to
all the Elodie RV measurements and the SB orbital elements were derived from Elodie alone.
The residuals of the RVs were calculated and
the uncertainty of the RVs was then chosen so that the $F_2$ estimator of the goodness-of-fit
is zero. According to \citet{GOF}, $F_2$ is derived from the formula

\begin{equation}
F_2= \left( \frac{9\nu}{2} \right)^{1/2} \left[ \left( \frac{\chi^2}{\nu} \right)^{1/3}+{\frac{2}{9 \nu}}-1 \right]
,\end{equation}

where $\nu$ is the number of degrees of freedom and $\chi^2$ is the weighted
sum of the squares of the differences between the predicted and the observed
RVs, normalised with respect to their uncertainties. It was thus found that the
uncertainty is 44~m~s$^{-1}$ for GJ~1069 and 14~m~s$^{-1}$ for GJ~554.
The final elements were then derived taking into account simultaneously 
the RVs from Elodie and those of CORAVEL. A systematic offset between the two sets
of measurements was derived with the SB solution. It is worth noticing that, 
although the CORAVEL RVs have weights much smaller than those of Elodie, they
slightly improve the accuracy of the solution terms.

\subsection{Presentation of the orbital elements}
\label{orbitPresentation}

We used the RV measurements to derive the SB orbital elements for all stars sufficiently observed. 
The number of these stars is 45; they are flagged ``O'' in Table~\ref{table:meanRV}. An SB orbit based
on our CORAVEL RV measurements has already been published for three of them: GJ~1064B, GJ~692.1, and GJ~765.2;
the orbits of the last two stars are even SB+VB orbits, since interferometric observations were also 
taken into account; the references of these three orbits are given in the footnotes of Table~\ref{table:meanRV}.
The SB orbital elements of the remaining 42 stars are listed in Table~\ref{tab:orbites}.
Since two stars are triple spectroscopic systems, this table contains 44 orbits.

The footnotes
of Table~\ref{tab:orbites} indicate that 23 of these orbits were previously published on the basis 
of a part of our measurements or from RVs measured with a different instrument than CORAVEL. The 21
other orbits refer to new SBs.

The orbits of eight G-type stars mentioned in \citetalias{DM91} are included in Table~\ref{tab:orbites}. 
\citetalias{DM91} provided a preliminary orbit for six of these stars, but the orbital
elements are significantly improved there, thanks to additional RV measurements.
The 44 SB orbits in Table~\ref{tab:orbites} are presented in Figs.~\ref{fig:orb1}, \ref{fig:orb2}, and \ref{fig:orb3}.

\section{Conclusion}
\label{conclusion}

We have drawn up a catalogue of 5413 RV measurements obtained with CORAVEL for 269 stars, 
261 K-type dwarfs, and eight G-type dwarfs of the solar neighbourhood. These measurements were 
used to detect the SBs on which were based the statistical investigations of \citetalias{Halbwachs03}. 
We calculated the elements of 44 SB orbits, corresponding to 42 stellar systems. Twenty-one
orbits, corresponding to 20 stellar systems, are the first orbits ever published for these stars.

All these data will be available through the VizieR service of the Centre de Donn\'ee astronomique de
Strasbourg (CDS). The SB orbits and the corresponding RV measurements will also be included in
the on-line SB9 catalogue \citep[http://sb9.astro.ulb.ac.be/, ][]{SB9}.

\begin{acknowledgements}
It is a pleasure to thank Salim Hillali for his contribution to the derivation of
the orbits. The referee, Roger Griffin, indicated some relevant references. 
We have benefited during the entire period of these observations from the support of the Swiss National
Foundation and Geneva University. We are particularly grateful to our technicians Bernard Tartarat, Emile
Ischi, and Charles Maire for their dedication to that experiment for more than 20 years.
We made use of Simbad, the database of the CDS and of the SB9 catalogue.
\end{acknowledgements}

%
%

   \bibliographystyle{aa}
   \bibliography{Halbwachs}

\begin{appendix} 
\section{Tables of mean RVs and of orbital elements}

\longtab[1]{
\begin{longtable}{rlrcrrrrrrrrlc}
\caption{\label{table:meanRV} Average RV and the variability status. A flag in the last column indicates that the
orbital elements are presented in Table~\ref{tab:orbites}; it is ``*'' when the
orbital elements are derived for the first time, and ``+'' otherwise.}\\
\hline\hline
\multicolumn{3}{c}{Identifications}& $B-V$ & $\overline{RV}$\phantom{O} & $\epsilon$ & $\sigma_{RV}$\phantom{O} &$E/I$& $P(\chi^2)$ & $N_1$ & $N_{tot}$ & $\Delta T$ & spect. & Table \\
\multicolumn{1}{c}{GJ} &\multicolumn{2}{c}{HD/BD/HIP/AG}&        &km~s$^{-1}$&km~s$^{-1}$&km~s$^{-1}$&                       &       &          & & days     & status & \ref{tab:orbites} \\
\hline
\endfirsthead
\caption{Continued.}\\
\hline\hline
\multicolumn{3}{c}{Identifications}& $B-V$ & $\overline{RV}$\phantom{O} & $\epsilon$ & $\sigma_{RV}$\phantom{O} &$E/I$& $P(\chi^2)$ & $N_1$ & $N_{tot}$ & $\Delta T$ & spect.& Table \\
\multicolumn{1}{c}{GJ} &\multicolumn{2}{c}{HD/BD/HIP/AG}&        &km~s$^{-1}$&km~s$^{-1}$&km~s$^{-1}$&                       &       &          & & days     & status & \ref{tab:orbites}\\
\hline
\endhead
\hline
\endfoot
   5\phantom{.2A}&HD &    166   &0.75&   $-$6.57& 0.07&  0.33&  1.13& 0.195& 21&  21& 5410&CST & \\
  27\phantom{.2A}&HD &   3651   &0.85&  $-$32.99& 0.06&  0.32&  1.11& 0.198& 25&  25& 4043&CST & \\
  28\phantom{.2A}&HD &   3765   &0.93&  $-$63.28& 0.02&  0.25&  0.82& 1.000&183& 183& 8037&CST & \\
  30\phantom{.2A}&BD &+33   99  &1.13&  $-$36.29& 0.10&  0.31&  0.94& 0.540& 10&  10& 2884&CST & \\
  31.4\phantom{A}&HD &   4256   &0.99&    9.40& 0.09&  0.24&  0.77& 0.838& 12&  12& 3262&CST & \\
  33\phantom{.2A}&HD &   4628   &0.88&  $-$10.33& 0.08&  0.34&  1.11& 0.242& 17&  17& 5132&CST & \\
  39\phantom{.2A}&HD &   4913   &1.21&    6.56& 0.10&  0.32&  0.90& 0.666& 14&  14& 3266&CST & \\
  44\phantom{.2A}&BD &$-$02  129  &0.83&  $-$48.00& 0.11&  0.26&  0.72& 0.863& 10&  10& 3261&CST & \\
  50\phantom{.2A}&BD &$-$10  216  &1.25&   $-$1.16& 0.09&  0.52&  1.19& 0.075& 33&  33& 5588&CST & \\
  52\phantom{.2A}&BD &+63  137  &1.30&    2.73& 0.13&  0.41&  1.06& 0.350& 10&  10& 2994&CST & \\
  53.1A&HD &   6660   &1.12&    6.58& 0.08&  0.28&  0.87& 0.719& 15&  15& 4838&CST & \\
  53.2\phantom{A}&BD &+16  120  &1.27&  $-$57.42& 0.14&  0.39&  0.93& 0.579&  9&   9& 2623&CST & \\
  55.2\phantom{A}&HIP&   5957   &1.36&  $-$23.66& 0.13&  0.28&  0.70& 0.868&  9&   9& 2629&CST & \\
  56\phantom{.2A}&HD &   7808   &1.00&   $-$7.31& 0.13&  0.30&  0.71& 0.896& 11&  11& 2968&CST & \\
  56.3A&HD &   7895   &0.82&   16.37& 0.09&  0.24&  0.76& 0.870& 13&  13& 3603&CST & \\
  56.4\phantom{A}&BD &+79   38  &1.29&  $-$20.61& 0.13&  0.37&  0.94& 0.529&  9&   9& 2626&CST & \\
  56.5\phantom{A}&HD &   7924   &0.82&  $-$22.63& 0.10&  0.26&  0.81& 0.727&  9&   9& 2621&CST & \\
  57.1A&HD &   8389   &0.91&   34.55& 0.08&  0.25&  0.81& 0.811& 14&  14& 2955&CST & \\
  57.1B&   &          &1.38&   34.78& 0.12&  0.43&  0.95& 0.573& 13&  13& 2955&CST & \\
  58.2\phantom{A}&HD &   8997   &0.96&   21.35& 0.10& 27.50& 60.04& 0.000& 31&  54& 2662&SB2O&+\\
1040\phantom{.2A}&HIP&   7655   &1.20&   32.30& 0.15&  0.52&  1.23& 0.134& 12&  12& 5229&CST & \\
  67\phantom{.2A}&HD &  10307   &0.63&    3.33& 0.05&  1.59&  4.90& 0.000& 32&  32& 7416&SB1O&+\\
  68\phantom{.2A}&HD &  10476   &0.84&  $-$33.67& 0.08&  0.24&  0.80& 0.844& 15&  15& 5846&CST & \\
  69\phantom{.2A}&HD &  10436   &1.22&  $-$50.94& 0.10&  0.31&  0.88& 0.680& 13&  13& 2958&CST & \\
  74\phantom{.2A}&HD &  10853   &1.04&   21.56& 0.10&  0.36&  1.00& 0.462& 13&  13& 3249&CST & \\
  75\phantom{.2A}&HD &  10780   &0.81&    2.82& 0.08&  0.28&  0.90& 0.642& 14&  14& 4711&CST & \\
  90\phantom{.2A}&HD &  13579   &0.92&  $-$12.58& 0.10&  0.17&  0.58& 0.950&  9&   9& 2621&CST & \\
  91.1A&HD &  13959   &1.10&   $-$0.17& 0.11&  0.54&  1.42& 0.004& 23&  23& 7428&SB1?& \\
  91.2A&HD &  14001   &1.02&    2.89& 0.12&  0.24&  0.65& 0.922& 10&  10& 5427&CST & \\
  92.1\phantom{A}&HD &  14039   &0.92&    9.02& 0.05& 11.14& 34.47& 0.000& 52&  52& 5588&SB1O&*\\
  98\phantom{.2}A&HD &  15285   &1.39&    6.57& 0.14&  4.71&  3.95& 0.000& 27&  54& 7410&SB2 & \\
 105\phantom{.2}A&HD &  16160   &0.97&   25.53& 0.06&  0.28&  0.95& 0.603& 24&  24& 6837&CST & \\
 106\phantom{.2A}&HD &  16909   &1.07&   31.28& 0.06&  4.41& 13.32& 0.000& 23&  23& 3604&SB1O&+\\
 112\phantom{.2A}&HD &  17190   &0.84&   14.00& 0.08&  0.28&  0.92& 0.631& 15&  15& 2891&CST & \\
 112.1\phantom{A}&HD &  17230   &1.28&   11.01& 0.10&  0.32&  0.91& 0.623& 12&  12& 3303&CST & \\
 113\phantom{.2A}&HD &  17382   &0.82&    9.11& 0.06&  2.07&  6.47& 0.000& 26&  26& 5592&SB1O&+\\
 114\phantom{.2A}&HD &  17660   &1.27&  $-$29.02& 0.12&  0.28&  0.81& 0.731&  9&   9& 2615&CST & \\
 117\phantom{.2A}&HD &  17925   &0.87&   17.92& 0.06&  0.30&  1.02& 0.422& 24&  24& 4851&CST & \\
 118.2A&HD &  18143   &0.93&   32.06& 0.11&  0.17&  0.54& 0.959&  8&   8& 2950&CST & \\
 120.1C&HD &  18445   &0.63&   49.90& 0.07&  0.93&  2.85& 0.000& 32&  32& 5929&SB1O&*\\
 138.1A&HD &  20727   &0.63&    7.51& 0.08&  4.07& 12.17& 0.000& 31&  31& 3982&SB1O&+\\
 141\phantom{.2A}&HD &  21197   &1.16&  $-$13.22& 0.10&  0.29&  0.94& 0.546& 10&  10& 2959&CST & \\
 142\phantom{.2A}&HD &  21531   &1.34&   34.35& 0.58&  1.00&  3.04& 0.000&  3&   3& 2039&SB1 & \\
 144\phantom{.2A}&HD &  22049   &0.88&   16.31& 0.06&  0.28&  1.02& 0.407& 21&  21& 3989&CST & \\
 150.2\phantom{A}&HD &  23140   &0.86&   22.24& 0.04&  3.94& 12.38& 0.000& 38&  38& 5585&SB1O&*\\
 153\phantom{.2}A&HD &  23189   &1.30&    2.58& 0.14&  0.20&  0.48& 0.986&  9&   9& 3984&CST & \\
1063\phantom{.2A}&BD &+11  514  &1.18&   83.73& 0.13&  0.10&  0.28& 0.999&  8&   8& 2889&CST & \\
1064\phantom{.2}A&HD &  23439   &0.78&   50.66& 0.12&  0.40&  1.03& 0.388& 11&  11& 6113&CST & \\
1064\phantom{.2}B&AG &+41  397  &0.90&   51.19& 0.06&  7.08& 18.94& 0.000& 29&  29& 3344&SB1O
\footnote{Our measurements were taken into account in the SB orbit of \citet{Tokovinin94}.} & \\
 155.2\phantom{A}&HD &  24238   &0.83&   38.95& 0.12&  0.23&  0.70& 0.842&  8&   8& 2265&CST & \\
 156.2\phantom{A}&HD &  24451   &1.15&   17.69& 0.10&  0.20&  0.62& 0.945& 10&  10& 3245&CST & \\
 157\phantom{.2}A&HD &  24916   &1.12&    3.64& 0.11&  0.29&  0.85& 0.687&  9&   9& 3405&CST & \\
 158\phantom{.2A}&HD &  25329   &0.88&  $-$25.89& 0.13&  0.53&  1.16& 0.173& 17&  17& 5564&CST & \\
 161\phantom{.2A}&HD &  25665   &0.91&  $-$13.57& 0.10&  0.30&  0.94& 0.552& 10&  10& 3245&CST & \\
 165.1\phantom{A}&HD &  26581   &1.00&   23.99& 0.12&  0.25&  0.74& 0.804&  8&   8& 3136&CST & \\
 165.2\phantom{A}&HD &  26794   &0.97&   56.51& 0.11&  0.36&  1.00& 0.448& 10&  10& 3410&CST & \\
1069\phantom{.2A}&BD &$-$04  782  &1.22&   25.23& 0.02&  0.85&  2.15& 0.000& 25&  25& 5223&SB1O&*\\
 166\phantom{.2}A&HD &  26965   &0.82&  $-$42.52& 0.07&  0.35&  1.16& 0.131& 22&  22& 5913&CST & \\
 168\phantom{.2A}&BD &+47  977  &1.17&  $-$78.60& 0.13&  0.37&  0.97& 0.480&  8&   8& 3248&CST & \\
 171\phantom{.2A}&HD & 237287   &0.89&   45.06& 0.09&  4.86& 15.09& 0.000& 26&  26& 5517&SB1O&+\\
 171.2A&HD & 283750   &1.12&   35.51& 0.04&  7.70& 22.48& 0.000& 54&  54& 2492&SB1O&+\\
 173.1A&HD & 286955   &1.02&  $-$27.57& 0.08&  3.23&  8.65& 0.000& 24&  24& 3306&SB1O&*\\
 174\phantom{.2A}&HD &  29697   &1.19&    1.01& 0.06&  0.32&  1.00& 0.482& 30&  30& 5840&CST & \\
 176.2\phantom{A}&HD &  29883   &0.92&   17.67& 0.10&  0.24&  0.74& 0.845& 10&  10& 3245&CST & \\
2035\phantom{.2A}&HD &  30973   &1.01&   26.24& 0.13&  0.29&  0.80& 0.721&  8&   8& 2891&CST & \\
 183\phantom{.2A}&HD &  32147   &1.06&   21.45& 0.07&  0.24&  0.82& 0.839& 18&  18& 4164&CST & \\
 200\phantom{.2}A&HD &  34673   &1.04&   87.84& 0.09&  0.22&  0.71& 0.886& 11&  11& 2628&CST & \\
1076\phantom{.2A}&BD &+54  886  &1.30&   53.71& 0.14&  0.57&  1.55& 0.002& 17&  17& 5409&SB1?& \\
 204\phantom{.2A}&HD &  36003   &1.10&  $-$55.72& 0.11&  0.36&  1.16& 0.211& 10&  10& 2628&CST & \\
 211\phantom{.2A}&HD &  37394   &0.84&    1.35& 0.09&  0.35&  1.15& 0.186& 14&  14& 4802&CST & \\
 217\phantom{.2A}&HD &  38230   &0.83&  $-$29.17& 0.09&  0.27&  0.84& 0.768& 14&  14& 3140&CST & \\
 221\phantom{.2A}&BD &$-$06 1339  &1.32&   22.87& 0.18&  0.50&  1.10& 0.302&  8&   8& 3425&CST & \\
 223\phantom{.2A}&HD &  39715   &1.01&  $-$33.80& 0.13&  0.33&  0.92& 0.560&  8&   8& 3399&CST & \\
 226.2\phantom{A}&HIP&  29067   &1.25&   $-$1.90& 0.13&  0.20&  0.52& 0.977&  9&   9& 3256&CST & \\
 227\phantom{.2A}&HD &  41593   &0.81&   $-$9.75& 0.09&  0.24&  0.78& 0.816& 11&  11& 4815&CST & \\
 233\phantom{.2A}&HD &  45088   &0.94&   $-$8.92& 0.08& 42.25&139.71& 0.000& 41&  46& 5603&SB2O&+\\
 241\phantom{.2A}&HD &  47752   &1.02&  $-$44.29& 0.10&  0.12&  0.36& 0.999& 11&  11& 3072&CST & \\
 249\phantom{.2A}&HD &  49601   &1.24&   19.44& 0.14&  0.46&  1.27& 0.107& 11&  11& 2943&CST & \\
 250\phantom{.2}A&HD &  50281   &1.05&   $-$7.20& 0.11&  0.39&  1.29& 0.076& 12&  12& 3238&CST & \\
 254\phantom{.2A}&HD & 266611   &1.36&  $-$14.95& 0.15&  0.40&  0.90& 0.603&  9&   9& 4410&CST & \\
 256\phantom{.2A}&HD &  51849   &1.13&   $-$5.84& 0.39&  1.47&  3.45& 0.000& 14&  14& 5416&SB1 & \\
 257.1\phantom{A}&HD &  51866   &0.99&  $-$21.51& 0.11&  0.28&  0.86& 0.664&  9&   9& 3071&CST & \\
1094\phantom{.2A}&HD &  52919   &1.08&  $-$30.81& 0.12&  0.30&  0.85& 0.688&  9&   9& 2561&CST & \\
 267\phantom{.2A}&HD &  54359   &0.96&   26.44& 0.12&  0.33&  0.93& 0.553&  9&   9& 4611&CST & \\
 273.1\phantom{A}&BD &+32 1561  &0.95&   $-$3.98& 0.09&  0.24&  0.82& 0.776& 12&  12& 2601&CST & \\
 276\phantom{.2A}&HD &  59582   &1.10&   66.03& 0.12&  0.26&  0.71& 0.860&  9&   9& 2615&CST & \\
 282\phantom{.2}A&HD &  61606   &0.96&  $-$18.21& 0.09&  0.31&  1.02& 0.418& 11&  11& 4760&CST & \\
 282\phantom{.2}B&BD &$-$03 2002  &1.33&  $-$19.02& 0.14&  0.43&  1.10& 0.289& 10&  10& 4760&CST & \\
 292.1\phantom{A}&HD &  64468   &0.95&  $-$15.91& 0.10&  3.97& 12.55& 0.000& 24&  24& 6124&SB1O&*\\
 293.1A&HD &  65277   &1.04&   $-$4.34& 0.13&  0.37&  1.10& 0.294&  8&   8& 2958&CST & \\
 295.1\phantom{A}&BD &+14 1802  &1.28&   20.35& 0.19&  0.55&  1.33& 0.103&  8&   8& 3984&CST & \\
 301.1\phantom{A}&BD &+31 1781  &1.14&   13.47& 0.10&  0.31&  0.97& 0.499& 10&  10& 2617&CST & \\
 313\phantom{.2A}&HD &  73583   &1.12&   20.22& 0.13&  0.22&  0.61& 0.925&  8&   8& 3277&CST & \\
 315\phantom{.2A}&HD &  73667   &0.82&  $-$12.03& 0.11&  0.25&  0.75& 0.808&  9&   9& 3591&CST & \\
1113\phantom{.2A}&HD &  73554   &1.08&   54.61& 0.13&  0.19&  0.53& 0.962&  8&   8& 3303&CST & \\
 321\phantom{.2A}&HD &  74377   &0.94&  $-$23.63& 0.15&  0.41&  1.22& 0.182&  7&   7& 3593&CST & \\
 325\phantom{.2}A&HD &  75632   &1.39&   43.52& 0.97&  2.92&  3.27& 0.000&  9&  18& 5516&SB2 & \\
 330.1\phantom{A}&BD &+21 1949  &1.11&  $-$50.36& 0.14&  0.14&  0.35& 0.997&  8&   8& 3936&CST & \\
 334.1\phantom{A}&BD &+73  447  &1.26&  $-$27.09& 0.16&  0.46&  1.03& 0.409&  8&   8& 5056&CST & \\
 337\phantom{.2}A&HD &  79096   &0.73&   49.99& 0.07&  8.50& 11.27& 0.000& 56& 101& 5391&SB2O&+\\
 338.1A&BD &+77  361  &1.38&  $-$11.34& 0.19&  0.56&  1.25& 0.149&  9&   9& 4050&CST & \\
 339\phantom{.2A}&HD &  79555   &1.02&    6.74& 0.22&  0.70&  2.18& 0.000& 10&  10& 3991&SB1 & \\
 340\phantom{.2}A&HD &  79969   &1.02&  $-$20.52& 0.06&  0.23&  0.72& 0.983& 27&  27& 5796&CST & \\
 340.2\phantom{A}&HD &  80367   &0.87&   50.95& 0.11&  0.31&  0.93& 0.549&  9&   9& 4091&CST & \\
 340.3\phantom{A}&HD &  80632   &1.17&   36.55& 0.13&  0.30&  0.77& 0.800&  9&   9& 3703&CST & \\
1124\phantom{.2A}&HD &  80715   &0.99&   $-$4.23& 0.08& 48.85& 90.03& 0.000& 48& 100& 1462&SB2O&*\\
 341.1\phantom{A}&BD &+81  297  &1.23&  $-$18.02& 0.17&  0.44&  1.23& 0.183&  7&   7& 3598&CST & \\
 342\phantom{.2A}&HD &  80768   &1.19&   $-$6.32& 0.10&  0.27&  0.69& 0.938& 14&  14& 3613&CST & \\
 343.1\phantom{A}&BD &+40 2208  &1.32&  $-$32.13& 0.13& 36.78& 53.52& 0.000& 21&  44& 1129&SB2O&*\\
 349\phantom{.2A}&HD &  82106   &1.00&   29.97& 0.10&  0.19&  0.60& 0.955& 10&  10& 3694&CST & \\
 355\phantom{.2A}&HD &  82558   &0.91&    7.58& 0.27&  0.81&  1.17& 0.211&  9&   9& 2140&CST & \\
 365\phantom{.2A}&HD &  84035   &1.15&  $-$12.30& 0.11&  0.32&  1.01& 0.421&  8&   8& 3598&CST & \\
 378.1\phantom{A}&HD &  86856   &1.07&   30.35& 0.11&  0.32&  0.94& 0.551&  9&   9& 5073&CST & \\
 379\phantom{.2}A&BD &+75  403  &1.40&  $-$55.35& 0.20&  0.50&  1.12& 0.306&  6&   6& 3563&CST & \\
 380\phantom{.2A}&HD &  88230   &1.36&  $-$26.21& 0.09&  0.22&  0.70& 0.898& 11&  11& 3568&CST & \\
 388.2\phantom{A}&HD &  89707   &0.63&   82.70& 0.09&  2.34&  6.26& 0.000& 64&  64& 6144&SB1O&+\\
 389.1\phantom{A}&BD &$-$09 3063  &1.23&   $-$3.97& 0.30&  0.90&  2.20& 0.000&  9&   9& 5145&SB1
\footnote{The RV is constant when the measurement of 13 April 1987 is discarded.} & \\
 394\phantom{.2A}&HD & 237903   &1.36&    8.84& 0.15&  0.23&  0.63& 0.856&  6&   6& 3663&CST & \\
 396\phantom{.2A}&HD &  90343   &0.82&    9.80& 0.10&  0.30&  0.95& 0.520&  9&   9& 3691&CST & \\
 397\phantom{.2A}&BD &+46 1635  &1.33&   20.62& 0.10&  0.35&  1.02& 0.434& 13&  13& 3979&CST & \\
 402.1\phantom{A}&BD &+00 2709  &0.90&  $-$48.33& 0.13&  0.30&  0.72& 0.873& 10&  10& 4769&CST & \\
1139\phantom{.2A}&BD &+76  404  &1.10&  $-$25.82& 0.15&  0.43&  1.02& 0.402&  8&   8& 3037&CST & \\
 414\phantom{.2}A&HD &  97101   &1.35&  $-$16.63& 0.11&  0.27&  0.75& 0.846& 11&  11& 3984&CST & \\
 418\phantom{.2A}&HD &  97503   &1.18&   16.41& 0.11&  0.31&  0.89& 0.626& 10&  10& 3988&CST & \\
 420\phantom{.2}A&HD &  97584   &1.04&    9.12& 0.12&  0.31&  0.97& 0.467&  7&   7& 3602&CST & \\
 426\phantom{.2}A&HD &  98736   &0.89&   $-$3.37& 0.11&  0.23&  0.74& 0.798&  8&   8& 3960&CST & \\
 429\phantom{.2}A&HD &  99491   &0.80&    4.14& 0.09&  0.26&  0.85& 0.727& 13&  13& 4754&CST & \\
 429\phantom{.2}B&HD &  99492   &1.00&    3.59& 0.13&  0.29&  0.96& 0.469&  6&   6& 3607&CST & \\
 435.1\phantom{A}&BD &+05 2529  &1.24&   19.48& 0.45&  2.48&  3.43& 0.000& 31&  57& 5759&SB2 & \\
 439\phantom{.2A}&BD &+31 2290  &1.13&   29.88& 0.12&  0.17&  0.50& 0.973&  8&   8& 3229&CST & \\
 441\phantom{.2A}&BD &+72  545  &1.17&  $-$17.13& 0.10&  7.08&  8.75& 0.000& 29&  56& 5453&SB2O&*\\
 444\phantom{.2}A&HD & 102392   &1.12&   18.99& 0.11&  0.28&  0.77& 0.814& 10&  10& 3689&CST & \\
 454\phantom{.2A}&HD & 104304   &0.76&    0.28& 0.08&  0.28&  0.96& 0.523& 12&  12& 4774&CST & \\
 469.1\phantom{A}&HD & 108754   &0.63&    0.40& 0.07&  5.11& 14.71& 0.000& 39&  39& 3692&SB1O&+\\
1160\phantom{.2A}&HD & 109011   &0.93&  $-$10.45& 0.15&  6.99& 11.73& 0.000& 35&  70& 5475&SB2O&*\\
 479.1\phantom{A}&HD & 110010   &0.63&  $-$18.31& 0.05&  2.12&  6.30& 0.000& 36&  36& 6141&SB1O&+\\
 481\phantom{.2A}&HD & 110315   &1.12&   24.90& 0.10&  0.24&  0.71& 0.873& 10&  10& 3292&CST & \\
 483\phantom{.2A}&HD & 110833   &0.94&    9.52& 0.15&  0.79&  2.63& 0.000& 48&  48& 6129&SB1O&*\\
 488.2\phantom{A}&BD &$-$05 3596  &1.34&  $-$12.84& 0.14&  0.49&  1.15& 0.224& 12&  12& 3297&CST & \\
 489\phantom{.2A}&HD & 112575   &1.12&   $-$7.77& 0.06&  1.80&  5.02& 0.000& 41&  41& 5892&SB1O&*\\
 491\phantom{.2}A&HD & 112758   &0.79&    3.95& 0.01&  1.38&  4.02& 0.000& 32&  32& 7413&SB1O&*\\
 498\phantom{.2A}&HD & 113827   &1.17&   $-$6.51& 0.13&  0.32&  0.83& 0.707&  9&   9& 3568&CST & \\
 505\phantom{.2}A&HD & 115404   &0.92&    7.51& 0.08&  0.25&  0.81& 0.817& 14&  14& 3244&CST & \\
 509\phantom{.2}A&HD & 116495   &1.33&  $-$39.18& 0.12&  0.44&  1.07& 0.338& 13&  13& 3981&CST & \\
 511\phantom{.2A}&HD & 116858   &0.93&  $-$12.61& 0.13&  0.43&  1.08& 0.325& 11&  11& 2956&CST & \\
2102\phantom{.2A}&HD & 117936   &0.80&   $-$6.00& 0.15&  0.41&  1.26& 0.136&  8&   8& 2958&CST & \\
 517\phantom{.2A}&HD & 118100   &1.18&  $-$22.47& 0.11&  0.48&  1.15& 0.186& 18&  18& 4012&CST & \\
1176\phantom{.2A}&HD & 119291   &1.19&  $-$43.17& 0.13&  0.23&  0.62& 0.917&  8&   8& 2903&CST & \\
 521.1\phantom{A}&HD & 118926   &1.39&   27.27& 0.20&  0.64&  1.33& 0.071& 10&  10& 4189&CST & \\
 523\phantom{.2A}&BD &+39 2675  &1.10&    1.24& 0.13&  0.34&  0.93& 0.542&  8&   8& 2905&CST & \\
 528\phantom{.2}A&HD & 120476   &1.12&  $-$20.18& 0.15&  0.36&  1.04& 0.375&  6&   6& 5577&CST & \\
 529\phantom{.2A}&HD & 120467   &1.27&  $-$38.34& 0.13&  0.27&  0.78& 0.732&  7&   7& 4008&CST & \\
 542.2\phantom{A}&HD & 125354   &1.30&   11.55& 0.51&  2.44&  5.76& 0.000& 23&  23& 6151&SB1O&*\\
 544\phantom{.2}A&HD & 125455   &0.84&   $-$9.93& 0.10&  0.21&  0.65& 0.929& 10&  10& 3282&CST & \\
 546\phantom{.2A}&BD &+30 2512  &1.26&  $-$37.18& 0.08&  0.30&  0.89& 0.699& 17&  17& 3688&CST & \\
 554\phantom{.2A}&HD & 127506   &1.13&  $-$18.69& 0.07&  0.26&  0.76& 0.946& 24&  24& 5893&CSTO
\footnote{The SB nature of the star was inferred from Elodie observations.}&* \\
 556\phantom{.2A}&HD & 128165   &0.99&   11.46& 0.08&  0.22&  0.76& 0.880& 14&  14& 4040&CST & \\
 561\phantom{.2A}&BD &+27 2411  &0.80&  $-$77.89& 0.12&  0.35&  0.88& 0.676& 11&  11& 2904&CST & \\
 562\phantom{.2A}&BD &+17 2785  &1.26&   42.47& 0.07&  0.35&  0.93& 0.720& 34&  34& 5721&CST & \\
 563.3\phantom{A}&HD & 130871   &0.97&  $-$32.32& 0.11&  0.25&  0.68& 0.916& 11&  11& 3291&CST & \\
 567\phantom{.2A}&HD & 131511   &0.84&  $-$31.42& 0.04& 11.19& 38.02& 0.000& 82&  82& 6669&SB1O&+\\
 569.1\phantom{A}&HD & 132142   &0.79&  $-$14.59& 0.09&  0.28&  0.87& 0.719& 14&  14& 4025&CST & \\
 570\phantom{.2}A&HD & 131977   &1.10&   26.73& 0.05&  0.29&  1.00& 0.465& 41&  41& 7364&CST & \\
 573\phantom{.2A}&HD & 132950   &1.04&   $-$1.73& 0.08&  0.36&  0.97& 0.563& 21&  21& 7272&CST & \\
 576\phantom{.2A}&BD &+06 2986  &1.30&  $-$84.79& 0.12&  0.43&  0.80& 0.806& 12&  12& 5092&CST & \\
 579\phantom{.2A}&BD &+25 2874  &1.36&  $-$69.43& 0.09&  0.52&  1.17& 0.100& 32&  32& 5500&CST & \\
1189\phantom{.2A}&BD &+24 2824  &1.06&  $-$53.64& 0.32&  1.33&  3.63& 0.000& 17&  17& 5161&SB1 & \\
 579.3\phantom{A}&HD & 134985   &0.78&  $-$61.49& 0.14&  0.45&  1.10& 0.290& 10&  10& 3689&CST & \\
 580\phantom{.2}A&HD & 135204   &0.77&  $-$69.54& 0.06&  0.23&  0.74& 0.962& 23&  23& 7337&CST & \\
1190\phantom{.2A}&BD &$-$03 3746  &1.13& $-$112.42& 0.13&  0.22&  0.50& 0.992& 11&  11& 4091&CST & \\
 583\phantom{.2A}&BD &$-$04 3873  &1.30&  $-$19.10& 0.15&  0.32&  0.77& 0.783&  8&   8& 2676&CST & \\
1192\phantom{.2A}&HD & 136834   &1.00&  $-$26.65& 0.07&  0.28&  0.85& 0.816& 22&  22& 4892&CST & \\
 586\phantom{.2}A&HD & 137763   &0.81&    7.47& 0.20& 21.20& 66.80& 0.000&110& 120& 6013&SB2O&+\\
 586\phantom{.2}B&HD & 137778   &0.90&    7.56& 0.07&  0.31&  0.98& 0.507& 18&  18& 6014&CST & \\
 591\phantom{.2A}&HD & 139323   &0.95&  $-$67.15& 0.09&  0.13&  0.47& 0.992& 10&  10& 5830&CST & \\
 593\phantom{.2}A&HD & 139341   &0.92&  $-$65.10& 1.00&  5.37& 11.08& 0.000& 29&  58& 6651&SB2 & \\
 610\phantom{.2A}&HD & 144253   &0.97&   36.76& 0.09& 19.12& 32.80& 0.000& 25&  47& 2973&SB2O&*\\
 612\phantom{.2A}&HD & 144872   &0.96&   23.43& 0.11&  0.35&  1.00& 0.452& 11&  11& 2991&CST & \\
 614\phantom{.2A}&HD & 145675   &0.87&  $-$13.88& 0.09&  0.26&  0.90& 0.614& 10&  10& 6291&CST & \\
 615.1A&HD & 145958   &0.77&   18.37& 0.11&  0.37&  1.01& 0.424& 11&  11& 5812&CST & \\
 615.1B&   &          &0.80&   18.50& 0.19&  0.63&  1.81& 0.000& 11&  11& 5812&SB1
\footnote{The RV is constant when the measurement of 22 May 1983 is discarded.} & \\
 621\phantom{.2A}&HD & 147776   &0.95&    7.08& 0.10&  0.32&  0.87& 0.699& 13&  13& 5225&CST & \\
 626\phantom{.2A}&HD & 148467   &1.22&  $-$36.54& 0.13&  0.13&  0.34& 0.998&  8&   8& 2994&CST & \\
 627\phantom{.2}A&HD & 148653   &0.86&  $-$31.13& 0.10&  0.28&  0.85& 0.704& 11&  11& 4849&CST & \\
 631\phantom{.2A}&HD & 149661   &0.81&  $-$12.95& 0.09&  0.18&  0.62& 0.952& 11&  11& 3803&CST & \\
 632.1\phantom{A}&HD & 149957   &1.20&  $-$11.16& 0.13&  0.27&  0.72& 0.824&  8&   8& 2963&CST & \\
 632.2A&BD &+76  614  &1.17&   $-$9.37& 0.12&  0.33&  0.82& 0.759& 11&  11& 4546&CST & \\
 637.1\phantom{A}&HD & 151541   &0.77&    9.57& 0.11&  0.20&  0.61& 0.934&  9&   9& 3339&CST & \\
 638\phantom{.2A}&HD & 151288   &1.37&  $-$32.07& 0.09&  0.26&  0.79& 0.855& 15&  15& 3619&CST & \\
 639\phantom{.2A}&HD & 151877   &0.82&    2.33& 0.12&  0.29&  0.86& 0.639&  8&   8& 2674&CST & \\
 640\phantom{.2A}&HD & 151995   &1.02&   $-$5.61& 0.14&  0.31&  0.87& 0.620&  7&   7& 3033&CST & \\
 649.1A&HD & 153557   &0.99&   $-$6.65& 0.09&  0.33&  1.02& 0.404& 15&  15& 4847&CST & \\
 649.1C&HD & 153525   &1.00&   $-$7.15& 0.09&  0.27&  0.83& 0.790& 15&  15& 4847&CST & \\
 650\phantom{.2A}&HD & 153631   &0.63&   83.21& 0.09&  4.10& 11.36& 0.000& 38&  38& 5246&SB1O&+\\
 653\phantom{.2A}&HD & 154363   &1.16&   34.02& 0.08&  0.30&  0.86& 0.779& 18&  18& 6307&CST & \\
 658\phantom{.2A}&HD & 155456   &0.87&  $-$59.79& 0.12&  0.11&  0.34& 0.997&  8&   8& 2655&CST & \\
 659\phantom{.2}A&HD & 155674   &1.16&    3.15& 0.13&  0.21&  0.55& 0.952&  8&   8& 2643&CST & \\
 659\phantom{.2}B&BD &+54 1862  &1.26&    2.23& 0.15&  0.32&  0.77& 0.768&  8&   8& 2643&CST & \\
 663\phantom{.2}A&HD & 155886   &0.86&    0.44& 0.10&  0.25&  0.81& 0.751& 10&  10& 3273&CST & \\
 663\phantom{.2}B&HD & 155885   &0.86&    0.10& 0.13&  0.40&  1.27& 0.113& 10&  10& 3273&CST & \\
 664\phantom{.2A}&HD & 156026   &1.16&   $-$0.09& 0.08&  0.28&  0.93& 0.620& 15&  15& 3694&CST & \\
 673\phantom{.2A}&HD & 157881   &1.36&  $-$23.89& 0.07&  0.26&  0.74& 0.958& 22&  22& 4477&CST & \\
 675\phantom{.2A}&HD & 158633   &0.76&  $-$38.56& 0.08&  0.28&  0.87& 0.727& 16&  16& 4821&CST & \\
 688\phantom{.2A}&HD & 160346   &0.96&   21.68& 0.05&  4.27& 14.16& 0.000& 38&  38& 4436&SB1O&+\\
 689\phantom{.2A}&HD & 160964   &1.10&  $-$24.41& 0.12&  0.36&  1.02& 0.410&  9&   9& 2561&CST & \\
 692.1\phantom{A}&HD & 161198   &0.77&   23.88&21.35&  6.23& 18.63& 0.000&105& 105& 4369&SB1O
\footnote{Our RV measurements were used by \citet{DTLGHM} to derive a combined spectroscopic and
interferometric orbit.} & \\
 697\phantom{.2A}&BD &+21 3245  &0.95&  $-$13.14& 0.17&  0.47&  1.29& 0.114&  8&   8& 2618&CST & \\
 698\phantom{.2A}&BD &+18 3497  &1.18&  $-$29.78& 0.17&  0.48&  1.24& 0.152&  8&   8& 2226&CST & \\
 700.2\phantom{A}&HD & 164922   &0.80&   20.14& 0.07&  0.22&  0.75& 0.911& 17&  17& 3717&CST & \\
 702\phantom{.2}A&HD & 165341   &0.86&   $-$9.73& 0.52&  2.16&  7.03& 0.000& 17&  17& 5911&SB1
\footnote{Combined spectroscopic and visual orbit by \citet{Berman32}, \citet{Batten76}, \citet{Batten84},
\citet{Heintz88}, \citet{Batten91}, and \citet{Pourbaix00}. \citet{Griffin83} and \citet{Griffin91}
pointed out the risk that the RV measurements are distorted due to contamination by the secondary component.}
 & \\
 706\phantom{.2A}&HD & 166620   &0.87&  $-$19.47& 0.07&  0.22&  0.74& 0.911& 16&  16& 4092&CST & \\
 715\phantom{.2A}&HD & 170493   &1.10&  $-$55.20& 0.09&  0.24&  0.76& 0.836& 11&  11& 2617&CST & \\
 718\phantom{.2A}&HD & 171314   &1.12&   38.25& 0.10&  0.20&  0.58& 0.971& 11&  11& 2612&CST & \\
 719\phantom{.2A}&HD & 234677   &1.22&  $-$25.53& 0.08& 19.30& 48.35& 0.000& 47&  88& 7096&SB2O&+\\
 722.1\phantom{A}&HD & 172393   &0.82&   32.24& 0.11&  0.28&  0.80& 0.764& 10&  10& 2620&CST & \\
 725.1\phantom{A}&HD & 173701   &0.84&  $-$45.55& 0.07&  0.31&  1.05& 0.352& 18&  18& 5204&CST & \\
 727\phantom{.2A}&HD & 174080   &1.08&   $-$7.17& 0.09&  0.17&  0.55& 0.986& 12&  12& 3020&CST & \\
 747.2\phantom{A}&BD &+33 3339  &1.25&    8.78& 0.64&  2.99&  3.34& 0.000& 22&  38& 5561&SB2 & \\
 753\phantom{.2A}&BD &+87  183  &1.06&   $-$5.74& 0.14&  0.29&  0.73& 0.843&  9&   9& 2569&CST & \\
 758\phantom{.2A}&HD & 182488   &0.81&  $-$21.56& 0.06&  0.20&  0.69& 0.971& 20&  20& 3026&CST & \\
1237\phantom{.2A}&HD & 183255   &0.93&  $-$64.95& 0.11& 10.65& 23.46& 0.000& 48&  96& 2259&SB2O&+\\
 761\phantom{.2}A&BD &+12 3917  &1.10&  $-$18.35& 0.13&  0.38&  1.01& 0.415&  9&   9& 2642&CST & \\
 762.1\phantom{A}&HD & 184467   &0.87&   11.63& 0.08&  7.14&  7.43& 0.000& 18&  36& 4105&SB2O&+\\
 764\phantom{.2A}&HD & 185144   &0.80&   26.64& 0.09&  0.35&  1.14& 0.202& 14&  14& 4426&CST & \\
 764.1A&HD & 184860   &1.01&   63.19& 0.23&  0.75&  2.08& 0.000& 11&  11& 5564&SB1 & \\
 764.1B&   &          &1.20&   64.68& 0.49&  0.85&  1.92& 0.027&  3&   3& 3698&CST & \\
 765.2\phantom{A}&HD & 186922   &0.88&   $-$5.21& 0.80&  5.81&  6.95& 0.000& 53& 103& 5516&SB2O
\footnote{Our RV measurements were used by \citet{Balega07} to derive a combined spectroscopic and
interferometric orbit.} & \\
 773.2\phantom{A}&HD & 189087   &0.80&  $-$29.68& 0.06&  0.24&  0.81& 0.873& 21&  21& 5557&CST & \\
 775\phantom{.2A}&HD & 190007   &1.12&  $-$30.51& 0.10&  0.32&  1.06& 0.350& 10&  10& 3272&CST & \\
 778\phantom{.2A}&HD & 190404   &0.82&   $-$2.55& 0.06&  0.30&  0.90& 0.777& 30&  30& 7337&CST & \\
 779.1\phantom{A}&HD & 190470   &0.91&   $-$7.33& 0.09&  0.12&  0.41& 0.999& 12&  12& 3023&CST & \\
 781.2\phantom{A}&HD & 191285   &1.12&  $-$18.55& 0.14&  0.44&  1.00& 0.449& 10&  10& 3008&CST & \\
 783.2A&HD & 191785   &0.85&  $-$49.52& 0.08&  0.29&  0.94& 0.583& 14&  14& 3369&CST & \\
 791.3\phantom{A}&BD &+33 3936  &1.13&  $-$27.05& 0.11&  0.15&  0.44& 0.995& 10&  10& 3008&CST & \\
 795\phantom{.2A}&HD & 196795   &1.22&  $-$40.86& 0.14&  2.27&  6.49& 0.000& 99&  99& 7660&SB1O&+\\
1255\phantom{.2}D&HD & 199476   &0.71&  $-$30.23& 0.13&  0.36&  1.02& 0.397&  8&   8& 1772&CST & \\
 808.2\phantom{A}&HD & 198550   &1.06&   $-$8.48& 0.10&  0.44&  1.42& 0.005& 21&  21& 5127&SB1?& \\
1259\phantom{.2A}&BD &+12 4499  &1.05&  $-$41.03& 0.06&  1.19&  3.57& 0.000& 29&  29& 5177&SB1O&*\\
 816.1A&HD & 200560   &0.97&  $-$14.11& 0.09&  0.21&  0.69& 0.920& 12&  12& 2961&CST & \\
 818\phantom{.2A}&HD & 200779   &1.22&  $-$66.81& 0.09&  0.31&  0.94& 0.584& 14&  14& 3340&CST & \\
 819\phantom{.2}A&HD & 200968   &0.90&  $-$32.72& 0.09&  0.15&  0.48& 0.996& 12&  12& 2967&CST & \\
 820\phantom{.2}A&HD & 201091   &1.17&  $-$65.82& 0.06&  0.34&  1.11& 0.179& 30&  30& 6226&CST & \\
 820\phantom{.2}B&HD & 201092   &1.37&  $-$64.67& 0.06&  0.30&  0.91& 0.725& 29&  29& 6226&CST & \\
 824\phantom{.2A}&HD & 202575   &1.02&  $-$18.24& 0.09&  0.27&  0.83& 0.760& 13&  13& 2964&CST & \\
 825.3\phantom{A}&HD & 202751   &0.99&  $-$27.65& 0.09&  0.20&  0.65& 0.957& 13&  13& 3349&CST & \\
 828.4\phantom{A}&HD & 204417   &0.86&  $-$14.63& 0.09&  0.33&  0.99& 0.476& 14&  14& 3639&CST & \\
 836.8\phantom{A}&BD &+40 4631  &1.34&    9.75& 0.13&  0.21&  0.53& 0.973&  9&   9& 2966&CST & \\
 838.1A&HD & 207491   &1.04&  $-$11.22& 0.10&  0.21&  0.66& 0.933& 11&  11& 3338&CST & \\
 838.2\phantom{A}&HD & 207795   &0.83&  $-$29.29& 0.10&  0.32&  0.94& 0.569& 11&  11& 3342&CST & \\
 840\phantom{.2A}&HD & 208313   &0.92&  $-$13.42& 0.07&  0.21&  0.72& 0.934& 16&  16& 2976&CST & \\
 850\phantom{.2A}&HD & 210667   &0.82&  $-$19.55& 0.07&  0.18&  0.64& 0.982& 17&  17& 2976&CST & \\
 851.4\phantom{A}&BD &+56 2737  &0.72&    2.45& 0.15& 11.38& 19.42& 0.000& 76& 131& 7440&SB2O&*\\
 854\phantom{.2A}&BD &+67 1424  &1.15&   $-$4.44& 0.11&  0.33&  0.91& 0.611& 10&  10& 2951&CST & \\
 857.1A&BD &+21 4747  &1.19&   $-$7.15& 0.10&  0.32&  0.96& 0.516& 11&  11& 4063&CST & \\
 867.1A&HD & 214615   &0.63&  $-$12.57& 0.11&  0.67&  1.76& 0.000& 35&  35& 7389&SB1O&*\\
 870\phantom{.2A}&BD &+42 4471  &1.11&  $-$32.64& 0.15&  6.17& 15.98& 0.000& 25&  25& 1392&SB1O&*\\
1272\phantom{.2}A&BD &+10 4812  &1.13&   $-$1.60& 0.12&  0.24&  0.63& 0.941& 10&  10& 2583&CST & \\
 871.2\phantom{A}&HD & 215704   &0.80&  $-$51.46& 0.10&  0.27&  0.83& 0.715& 10&  10& 2912&CST & \\
 886\phantom{.2A}&HD & 217580   &0.95&  $-$44.19& 0.04&  1.78&  5.75& 0.000& 25&  25& 2891&SB1O&+\\
 892\phantom{.2A}&HD & 219134   &1.00&  $-$18.58& 0.07&  0.26&  0.92& 0.633& 16&  16& 4057&CST & \\
 893.2B&HD & 219430   &1.05&  $-$24.29& 0.34&  0.05&  0.10& 0.919&  1&   1&    0&?   \\
 894.4\phantom{A}&HD & 220182   &0.80&    3.32& 0.06&  0.23&  0.78& 0.923& 23&  23& 4435&CST & \\
 894.5\phantom{A}&HD & 220339   &0.89&   33.77& 0.10&  0.22&  0.69& 0.898& 10&  10& 2888&CST & \\
 895.4\phantom{A}&HD & 221354   &0.83&  $-$25.02& 0.09&  0.14&  0.46& 0.993& 10&  10& 2886&CST & \\
 904.1A&HD & 222474   &1.11&  $-$19.24& 0.12&  0.44&  1.33& 0.046& 14&  14& 5135&CST & \\
 907.1\phantom{A}&BD &$-$13 6464  &1.26&   $-$9.04& 0.45&  0.45&  1.00& 9.999&  1&   1&    0&?   \\
 908.1\phantom{A}&BD &+29 5007  &1.26&   $-$3.92& 0.11&  0.27&  0.75& 0.831& 10&  10& 2887&CST & \\
 909\phantom{.2}A&HD & 223778   &0.98&    4.54& 0.37& 22.54& 29.97& 0.000&  7&  13& 2853&SB2O&+\\
 909.1\phantom{A}&HD & 223782   &1.08&  $-$29.26& 0.12&  0.29&  0.73& 0.874& 11&  11& 5048&CST & \\
\end{longtable}
}


\longtab[2]{
\begin{landscape}
\begin{longtable}{lrrcrrrcrcc}
\caption{\label{tab:orbites} Orbital elements of the SBs. The G--type stars with an orbit in \citetalias{DM91} are
included in this table when their elements are significantly improved thanks to additional observations; they are
indicated by ``(G)'' following the GJ identification}\\
\hline\hline
HD/BD & $P$ & $T_0$(JD) & $e$ &  $V_0$ & $\omega_1$ & $K_{1,2}$ &  $m_{1,2} \sin^3 i$ or  &  $a_{1,2} \sin i$ &  $N_1$ & $\sigma(O-C)$ \\
GJ & (d) & 2400000+ &  & (km s$^{-1}$) & ($^{\circ}$) & (km s$^{-1}$)  &   $f_1(m)$ (M$_\odot$) &  (Gm) & $N_2$ & (km s$^{-1}$) \\
\hline
\endfirsthead
\caption{continued.}\\
\hline\hline
HD/BD & $P$ & $T_0$(JD) & $e$ &  $V_0$ & $\omega_1$ & $K_{1,2}$ &  $m_{1,2} \sin^3 i$ or  &  $a_{1,2} \sin i$ &  $N_{1,2}$ & $\sigma(O-C)$ \\
GJ & (d) & 2400000+ &  & (km s$^{-1}$) & ($^{\rm o}$) & (km s$^{-1}$)  &   $f_1(m)$ (M$_\odot$) &  (Gm) & $N_0$ & (km s$^{-1}$) \\
\hline
\endhead
\hline
\endfoot
HD 8997
\footnote{First orbit by \citet{Griffin87}; other orbit by \citet{Scarfe88}.}
       &   10.98358  &46482.39  &0.0358 & 21.35 &192.86 &38.87 &0.3809   &  5.867 & 31       &0.74 \\
GJ 58.2       &    0.00012  &    0.17  &0.0036 &  0.10 &  5.44 & 0.17 &0.0056   &  0.025 &          &     \\    
              &             &          &       &       &       &46.21 &0.3203   &  6.976 & 23       &     \\
              &             &          &       &       &       & 0.36 &0.0042   &  0.054 &          &     \\
&&&&&&&&&&\\
HD 10307
\footnote{Preliminary orbit in \citetalias{DM91}; first orbit with $\Delta T$ longer than the period.}
      & 7206.       &43259.    &0.437  &  3.329&203.71 & 2.966&0.0142   &264.3   & 32       &0.24 \\
GJ 67 (G)     &   42.       &   75.    &0.024  &  0.050&  4.51 & 0.110&0.0017   & 10.5   &          &     \\
&&&&&&&&&&\\
HD 14039
\footnote{Triple system.}
      &   80.0342   &46773.64  &0.3325 &  9.021& 40.24 &14.488&0.02120  & 15.037 & 52       &0.35 \\
GJ 92.1\,a-b  &    0.0029   &    0.16  &0.0051 &  0.050&  0.99 & 0.079&0.00037  &  0.087 &          &     \\
&&&&&&&&&&\\
HD 14039      & 4570.       &48277.0   &0.4731 &       &288.68 & 3.454&0.01338  &191.27  & 52       &0.34 \\
GJ 92.1\,ab-c &  106.       &   27.3   &0.0246 &       &  3.48 & 0.100&0.00135  &  7.67  &          &     \\ 
&&&&&&&&&&\\
HD 16909
\footnote{First orbit by \citet{3GZ85}.}
      & 1227.77     &46638.50  &0.492  & 31.281&119.35 & 6.476&0.0228   & 95.18  & 23       &0.24 \\
GJ 106        &    3.35     &    4.66  &0.014  &  0.064&  1.60 & 0.096&0.0012   &  1.66  &          &     \\
&&&&&&&&&&\\
HD 17382
\footnote{Preliminary orbit by \citet{Latham02}.}
      & 5954.       &48024.2   &0.663  &  9.111&110.74 & 2.957&0.00671  &181.3   & 26       &0.25 \\
GJ 113        &  294.       &   25.2   &0.021  &  0.065&  3.59 & 0.080&0.00081  & 11.2   &          &     \\
&&&&&&&&&&\\
HD 18445\,C   &  553.89     &47489.9   &0.528  & 49.896& 74.7  & 1.50 &0.000119 &  9.72  & 32     &0.38 \\
GJ 120.1\,C (G)&    1.37     &    8.7   &0.067  &  0.070&  9.1  & 0.13 &0.000035 &  0.96  &          &     \\
&&&&&&&&&&\\
HD 20727
\footnote{We improve the orbit of  \citetalias{DM91}.}
      &  325.07     &46515.6   &0.499  &  7.507&228.1  & 7.53 &0.00939  & 29.17  & 31       &0.40 \\
GJ 138.1\,A (G)&    0.39     &    1.6   &0.020  &  0.080&  1.9  & 0.20 &0.00083  &  0.86  &          &     \\
&&&&&&&&&&\\
HD 23140
\footnote{First SB orbit; the star was already a VB with known orbital elements \citep{Wardat14}.}
      & 5121.       &48636.9   &0.275  & 22.238& 72.30 & 5.745&0.08967  &389.02  & 38       &0.23 \\
GJ 150.2      &   45.       &   25.7   &0.010  &  0.044&  2.16 & 0.056&0.00284  &  5.29  &          &     \\
&&&&&&&&&&\\
BD$-$04 782
\footnote{Orbit derived taking into account 15 Elodie measurements found in the Elodie Archive; the offset of the Elodie RVs is -0.193 km s$^{-1}$.}
    &  716.80     &49992.2   &0.090  & 25.226&242.9  & 1.318&0.000168 & 12.93  & 25$+$15  &0.42, 0.034 \\
GJ 1069       &    2.01     &   27.8   &0.018  &  0.023& 15.0  & 0.041&0.000016 &  0.42  &          &     \\
&&&&&&&&&&\\
HD 237287
\footnote{First orbit by \citet{Tokovinin94}; other orbit by \citet{Latham02}.}
     &  330.45     &46785.14  &0.649  & 45.057&122.58 & 7.91 &0.00748  & 27.35  & 26       &0.30 \\
GJ 171        &    0.26     &    2.63  &0.016  &  0.086&  1.36 & 0.25 &0.00082  &  1.00  &          &     \\
&&&&&&&&&&\\
HD 283750\,A
\footnote{First orbit by \citet{3GZ85}; other orbit by \citet{Tokovinin90}.}
  &    1.787992 &46999.42  &0.0022 & 35.505&323.77 &10.454&0.0002121&  0.2570& 54       &0.29 \\
GJ 171.2\,A   &    0.000004 &    0.74  &0.0052 &  0.040&149.67 & 0.056&0.0000034&  0.0014&         &      \\
&&&&&&&&&&\\
HD 286955\,A  &  610.02     &47474.27  &0.399  &$-$27.565&266.96 & 4.55 &0.00461  & 35.01  & 24       &0.36 \\
GJ 173.1\,A   &    1.02     &    5.96  &0.023  &  0.078&  4.75 & 0.13 &0.00042  &  1.08  &          &     \\
&&&&&&&&&&\\
HD 45088
\footnote{First orbit by \citet{GrifEm75}.}
      &    6.991849 &48991.973 &0.1475 & $-$8.917& 79.65 &56.86 &0.651    &  5.407 & 41       &0.51 \\
GJ 233        &    0.000016 &    0.021 &0.0019 &  0.080&  0.85 & 0.11 &0.011    &  0.011 &          &     \\
              &             &          &       &       &       &63.70 &0.581    &  6.06  &  5       &     \\
              &             &          &       &       &       & 1.40 &0.015    &  0.13  &          &     \\
&&&&&&&&&&\\
HD 64468      &  161.119    &46586.84  &0.273  &$-$15.91 &323.98 & 5.69 &0.00275  & 12.13  & 24       &0.42 \\
GJ 292.1      &    0.079    &    2.01  &0.021  &  0.10 &  5.36 & 0.13 &0.00020  &  0.29  &          &     \\
&&&&&&&&&&\\
HD 79096\,AB
\footnote{First orbit by \citet{Griffin1982}; combined VB+SB2 orbits by \citet{MaMcAlHar96} and by \citet{Pourbaix00}.}
  &  987.82     &47197.82  &0.4431 & 49.989&170.41 &11.52 &0.487    &140.29  & 56       &0.65 \\
GJ 337\,AB    &    1.06     &    3.04  &0.0097 &  0.070&  1.00 & 0.12 &0.018    &  1.63  &          &     \\
              &             &          &       &       &       &11.95 &0.469    &145.56  & 45       &     \\
              &             &          &       &       &       & 0.21 &0.016    &  2.70  &          &     \\
&&&&&&&&&&\\
HD 80715      &    3.804067 &47456.2383&0.0000 & $-$4.23 &  0.0  &71.18 &0.5691   &  3.7236& 48       &0.70 \\
GJ 1124       &    0.000008 &    0.0025&0.0032 &  0.08 &       & 0.14 &0.0025   &  0.0072&          &     \\ 
              &             &          &       &       &       &71.13 &0.5696   &  3.7208& 51       &     \\
              &             &          &       &       &       & 0.14 &0.0025   &  0.0075&          &     \\
&&&&&&&&&&\\
BD+40 2208    &    8.490697 &46494.695 &0.0993 &$-$32.13 &326.68 &56.45 &0.6412   &  6.559 & 21       &0.81 \\
GJ 343.1      &    0.000066 &    0.047 &0.0033 &  0.13 &  2.01 & 0.27 &0.0069   &  0.032 &          &     \\
              &             &          &       &       &       &57.17 &0.6332   &  6.642 & 23       &     \\
              &             &          &       &       &       & 0.27 &0.0069   &  0.032 &          &     \\
&&&&&&&&&&\\
HD 89707
\footnote{We improve the orbit of \citetalias{DM91}; high precision orbit by \citet{Sahlmann11}.}
     &  298.295    &48517.20  &0.946  & 82.723&108.9  & 4.98 &0.000131 &  6.63  & 64      &0.32 \\
GJ 388.2(G)   &    0.122    &     0.50  &0.012  &  0.051&  4.8  & 0.41 &0.000053 &  0.90  &          &     \\
&&&&&&&&&&\\
BD+72 545     &  632.56     &49608.41  &0.249  &$-$17.13 &278.33 &10.05 &0.274    & 84.67  & 29       &0.77 \\
GJ 441        &    0.57     &    4.93  &0.018  &  0.10 &  3.31 & 0.20 &0.013    &  1.73  &          &     \\
              &             &          &       &       &       &10.69 &0.258    & 90.03  & 27       &     \\
              &             &          &       &       &       & 0.22 &0.012    &  1.95  &          &     \\
&&&&&&&&&&\\
HD 108754
\footnote{We improve the CORAVEL orbit of \citet{JasMay88}.}
     &   25.93115  &45956.44  &0.1759 & 0.403 & 59.0  & 7.713&0.001179 &  2.707 & 39       &0.39 \\
GJ 469.1 (G)  &    0.00100  &    0.30  &0.0117 & 0.069 &  4.2  & 0.100&0.000046 &  0.036 &          &     \\
&&&&&&&&&&\\
HD 109011     & 1284.38     &47513.68  &0.501  &$-$10.45 &250.84 & 9.82 &0.443    &150.13  & 35       &1.20 \\
GJ 1160       &    2.34     &    7.58  &0.017  &  0.15 &  2.98 & 0.24 &0.035    &  4.07  &          &     \\
              &             &          &       &       &       &11.38 &0.382    &173.99  & 35       &     \\
              &             &          &       &       &       & 0.41 &0.029    &  6.61  &          &     \\
&&&&&&&&&&\\
HD 110010 
\footnote{We improve the preliminary orbit of \citetalias{DM91}; first orbit with $\Delta T$ longer than the period.}
    & 4118.       &46030.    &0.199  &$-$18.312&350.60 & 3.156&0.01265  &175.1   & 37       &0.27 \\
GJ 479.1 (G)  &   26.       &   63.    &0.020  &  0.046&  6.05 & 0.069&0.00085  &  4.0   &          &     \\
&&&&&&&&&&\\
HD 110833     &  270.33     &49797.0   &0.902  &  9.568&251.6  & 1.72 &0.0000115&  2.77  & 47       &0.24 \\
GJ 483        &    0.54     &    1.0   &0.044  &  0.047&  7.4  & 0.31 &0.0000028&  0.23  &          &     \\
&&&&&&&&&&\\
HD 112575     & 3572.1      &49872.95  &0.766  & $-$7.773&238.51 & 2.48 &0.00151  & 78.42  & 41       &0.34 \\
GJ 489        &   34.5      &    9.04  &0.021  &  0.063&  3.93 & 0.10 &0.00025  &  4.45  &          &     \\
&&&&&&&&&&\\
HD 112758\,A  &  103.226    &49672.0   &0.141  &  3.945&333.   & 1.850&0.0000659&  2.60  & 32       &0.33 \\
GJ 491\,A     &   0.042     &    5.7   &0.044  &  0.061& 20.   & 0.083&0.0000089&  0.12  &          &     \\
&&&&&&&&&&\\
HD 125354
\footnote{The period of the interferometric orbit by \citet{TokoMas15} is assumed.}
     & 6819.       &45990.    &0.306  & 11.55 & 67.1  & 3.96  &0.0379   &353.    & 23       &0.49 \\
GJ 542.2      &             &  117.    &0.039  &  0.11 &  7.6  & 0.18  &0.0054   & 17.    &          &     \\
&&&&&&&&&&\\
HD 127506
\footnote{The orbital elements were derived taking into account Elodie RV measurements; the offset of the Elodie RVs is $-$0.028 km s$^{-1}$.}
     & 2669.5      &52527.29  &0.7086 &$-$19.117&244.01 & 0.9051&0.0000720& 23.44  & 24+56    &0.24 \\
GJ 554        &   17.7      &    1.52  &0.0046 &  0.004&  0.45 & 0.0057&0.0000012&  0.21  &          &0.013\\
&&&&&&&&&&\\
HD 131511
\footnote{First orbit by \citet{Kamper81}; other orbit by \citet{BeavSal83};
high precision orbits by \citet{NMBFV02} and by \citet{KITS13,KITS16}.}
     &  125.3939   &44936.898 &0.5096 &$-$31.425&219.48 &18.913&0.05612  & 28.06  & 82       &0.31 \\
GJ 567        &    0.0020   &    0.082 &0.0026 &  0.036&  0.39 & 0.076&0.00074  &  0.12  &          &     \\
&&&&&&&&&&\\
HD 137763\,A
\footnote{First orbit by \citet{Tokovinin91}; we improve the orbit of \citet{DMACN92}.}
  &  889.813    &47967.519 &0.9733 &  7.47 &252.64 &36.42 &0.4716   &102.29  &110       &0.74 \\
GJ 586\,A     &    0.017    &    0.015 &0.0006 &  0.20 &  0.73 & 0.38 &0.0146   &  0.70  &          &     \\
              &             &          &       &       &       &52.90 &0.3247   &148.57  & 10       &     \\
              &             &          &       &       &       & 1.73 &0.0083   &  2.18  &          &     \\
&&&&&&&&&&\\
HD 144253     &  105.947    &48948.02  &0.1514 & 36.759& 85.28 &23.55 &0.7176   & 33.92  & 25       &0.56 \\
GJ 610        &    0.017    &    0.90  &0.0055 &  0.088&  2.04 & 0.13 &0.0118   &  0.19  &          &     \\
              &             &          &       &       &       &26.72 &0.6323   & 38.48  & 22       &     \\
              &             &          &       &       &       & 0.22 &0.0097   &  0.32  &          &     \\
&&&&&&&&&&\\
HD 153631
\footnote{We improve the orbit of \citet{DM88}.}
     &  386.64     &46161.5   &0.186  & 83.208& 46.1  & 6.080&0.00856  & 31.76  & 38       &0.32 \\
GJ 650 (G)    &    0.30     &    4.3   &0.018  &  0.088&  4.9  & 0.085&0.00037  &  0.46  &          &     \\
&&&&&&&&&&\\
HD 160346
\footnote{First orbit by \citet{Tokovinin91}; high precision orbit by \citet{KITS13}.}
     &   83.714    &46972.20  &0.210  & 21.680&145.5  & 5.644&0.001460 &  6.351 & 38       &0.26 \\
GJ 688        &    0.012    &    0.57  &0.012  &  0.045&  2.7  & 0.057&0.000046 &  0.067 &          &     \\
&&&&&&&&&&\\
HD 234677
\footnote{The BY Dra variable star; first orbit by \citet{Bopp73}; other orbit by \citet{VogtFek79}.}
     &    5.975114 &49996.375 &0.3050 &$-$25.533&229.43 &28.43 &0.06225  &  2.225 & 47       &0.69 \\
GJ 719        &    0.000007 &    0.012 &0.0032 &  0.075&  0.77 & 0.13 &0.00078  &  0.011 &          &     \\ 
              &             &          &       &       &       &31.90 &0.05548  &  2.496 & 41       &     \\
              &             &          &       &       &       & 0.19 &0.00067  &  0.015 &          &     \\
&&&&&&&&&&\\
HD 183255saga.edpsciences.org/publication/aa/login/login/
\footnote{First orbit by \citet{Tokovinin91}; high precision orbit by \citet{Kiefer16}.}
     &  166.83     &46415.94  &0.141  &$-$64.95 & 69.0  &13.91 &0.2252   & 31.60  & 48       &1.01 \\
GJ 1237       &    0.18     &    1.85  &0.012  &  0.11 &  4.5  & 0.18 &0.0074   &  0.41  &          &     \\
              &             &          &       &       &       &15.49 &0.2022   & 35.18  & 48       &     \\
              &             &          &       &       &       & 0.24 &0.0064   &  0.55  &          &     \\
&&&&&&&&&&\\
HD 184467
\footnote{First orbit by \citet{McClure83}; combined VB+SB2 orbit by \citet{Pourbaix00}.
A high precision VB+SB2 orbit was derived by \citet{Kiefer16}.}
     &  494.77     &46662.73  &0.339  & 11.632&177.49 & 9.44 &0.1583   & 60.41  & 18       &0.48 \\
GJ 762.1      &    0.50     &    2.18  &0.013  &  0.083&  2.31 & 0.17 &0.0068   &  1.13  &          &     \\
              &             &          &       &       &       & 9.90 &0.1510   & 63.35  & 18       &     \\
              &             &          &       &       &       & 0.18 &0.0065   &  1.19  &          &     \\
&&&&&&&&&&\\
HD 196795
\footnote{Triple system; first orbit by \citet{Duquennoy87}.} 
     &  918.59     &48082.70  &0.6872 &$-$40.860&116.5  & 3.394&0.001430 & 31.14  & 99       &0.41 \\
GJ 795\,a-b   &    0.56     &    1.80  &0.0122 &  0.045&  2.0  & 0.068&0.000109 &  0.79  &          &     \\
&&&&&&&&&&\\
HD 196795
\footnote{The period and the semi-amplitude were fixed to the results of the SB+VB solution derived by \citet{TokoLatham17}.} 
     &14128.       &50796.    &0.089  &       & 72.   & 2.66 &0.0273   &514.70  & 99       &0.41 \\
GJ 795 ab-c   &             &  555.    &0.035  &       & 14.   &      &0.0026   &  1.61  &          &     \\
&&&&&&&&&&\\
BD+12 4499    & 2173.3      &47551.2   &0.459  &$-$41.029&233.58 & 1.93 &0.00113  & 51.14  & 29       &0.28 \\
GJ 1259       &   36.4      &   56.1   &0.044  &  0.060&  8.33 & 0.10 &0.00021  &  3.20  &          &     \\
&&&&&&&&&&\\
BD+56 2737    &  650.19     &47572.66  &0.5971 &  2.447& 12.63 &15.56  &0.938    &111.61  & 74       &1.12 \\
GJ 851.4      &    0.26     &    0.91  &0.0077 &  0.105&  1.03 & 0.17  &0.039    &  1.44  &          &     \\
              &             &          &       &       &       &20.59  &0.709    &147.65  & 53       &     \\
              &             &          &       &       &       & 0.38  &0.027    &  2.90  &          &     \\
&&&&&&&&&&\\
HD 214615
\footnote{Preliminary orbit based on observations that do not completely cover the orbital period.}
     & 7725.       &45741.    &0.54   &$-$12.57 & 11.43 & 1.36 &0.00122  &122.    & 35       &0.43 \\
GJ 867.1A (G) &  952.       &  221.    &0.13   &  0.11 & 12.49 & 0.22 &0.00070  & 27.    &          &     \\
&&&&&&&&&&\\
BD+42 4471    &  374.57     &47098.21  &0.440  &$-$32.64 &343.33 & 8.03 &0.01455  & 37.11  & 25       &0.32 \\
GJ 870        &    0.62     &    1.92  &0.012  &  0.15 &  2.68 & 0.13 &0.00077  &  0.66  &          &     \\
&&&&&&&&&&\\
HD 217580
\footnote{First orbit by \citet{Tokovinin94}, based on CORAVEL observations; other orbit by \citet{Latham02}}
     &  453.90     &49162.2   &0.518  &$-$44.186&239.78 & 2.501&0.000426 & 13.36  & 25       &0.27 \\
GJ 886        &    1.78     &    8.5   &0.041  &  0.058&  3.90 & 0.097&0.000067 &  0.65  &          &     \\
&&&&&&&&&&\\
HD 223778
\footnote{First orbit by \citet{Christie34}.}
     &    7.753470 &47000.4492&0.0000 &  4.54 &  0.00 &46.74 &0.3609   &  4.983 &  7       &0.32 \\
GJ 909\,A     &    0.000065 &    0.0094&0.0061 &  0.37 & 60.   & 0.75 &0.0045   &  0.025 &          &     \\
              &             &          &       &       &       &48.95 &0.3446   &  5.219 &  6       &     \\
              &             &          &       &       &       & 0.93 &0.0041   &  0.031 &          &     \\
\hline
\end{longtable}
\end{landscape}
}

\end{appendix}

\end{document}